\documentclass[3p]{elsarticle}        
\usepackage{natbib}
\usepackage{amsfonts,amsmath,amssymb}
\usepackage{bm}
\usepackage{hyperref}

\begin{document}

\begin{frontmatter}

\title{Displacement Data Assimilation}

\author{W. Steven Rosenthal$^1$}
\author{Shankar C. Venkataramani$^2$}
\author{Arthur J. Mariano$^3$}
\author{Juan M. Restrepo$^4$\corref{mycorrespondingauthor}}

\address{$^1$ Pacific Northwest Laboratory, Richland WA USA  99354\\
$^2$ Department of Mathematics and Program in Applied Mathematics, University of Arizona, Tucson AZ USA 85721\\
$^3$ Rosenstiel School of Marine \& Atmospheric Science, University of Miami, Miami FL USA 33149\\
$^4$ Department of Mathematics, Oregon State University, Corvallis OR USA 97331
}


\cortext[mycorrespondingauthor]{Corresponding Author: }
\ead{restrepo@math.oregonstate.edu}


\begin{abstract}
We show that modifying a Bayesian  data assimilation scheme by incorporating kinematically-consistent displacement corrections produces a  scheme that is demonstrably better at estimating partially observed state vectors  in a 
setting where feature information important. While the displacement transformation is not tied to any particular assimilation scheme, here we implement it within an ensemble Kalman Filter and demonstrate its effectiveness in tracking stochastically perturbed vortices. 
\end{abstract}


\end{frontmatter}


\section{Bayesian Estimation and Displacement Assimilation}

Most sequential estimation strategies are Bayesian. In these we seek to find estimates of moments (or of the whole probability density function, cdf) of the state variable $X(t)$, a random vector that is time dependent, subject to (usually
discrete) observations of the state, $Y(t)$. The posterior probability is thus
\[
P(X|Y) \propto P(Y|X) P(X),
\]
where the first term on the right hand side is the likelihood, and is informed by observations, and $P(X)$ is the prior, which is informed by the model. The model is usually an evolution equation for $X$. Both the model and the observations involve  stochastic processes. The model ''error''  might be an explicit stochastic term representing uncertainties in the evolution equation or its initial or boundary data. The observations are usually stochastic because there are inherent errors in the measurements.
When the prior and the likelihood are Gaussian (or a product of Gaussians), minimizing the quadratic arguments of $P(Y|X)$ and $P(X)$ maximizes $P(X|Y)$. Moreover, it does so by taking into account the relative certainties in the model and in the measurements. The most familiar sequential estimation technique is the Kalman Filter (see \cite{Jazw70}). It is optimal in the sense that it minimizes the trace of the posterior variance, yielding an estimate of the mean of  $X(t)$ for some interval in $t$, and the associated  covariance. This optimality is achieved when the relationship between the observations and the state variable is linear and the noise is Gaussian, and the  state variable $X(t)$ has linear dynamics and remains Gaussian for all time. 

For mildly nonlinear/non-Gaussian problems, the extended Kalman Filter (see \cite{Jazw70}) and the ensemble Kalman Filter
(enKF) \cite{Even09} are alternatives, though not guaranteed  to converge.  Nevertheless, the enKF will be employed in this study, as it provides a useful framework for developing more targeted assimilation methods. enKF is a two-stage sequential estimation process for the mean and the variance of the posterior. 
There is a forecast, wherein the model is advanced from $t -\delta t$ to $t$  to propose  an ensemble of  states. This is followed by an analysis stage.
if observations are available at that time $t$, these  are individually assimilated into each ensemble member using covariance information from the whole ensemble.   More details on the enKF filtering scheme appear in  \ref{enkf}.

Most data assimilation schemes are variance-minimizing; they give an appropriately weighted average the predictions of the model and the noisy observations \cite{wunschbook}.  These methods thus tend to decrease sharp gradients, consequently smearing or obliterating ``features" (vortices, shock-fronts, etc.)
in the state variable being estimated. If tracking the features are critical, a purely variance-minimizing methodology will produce estimates that might not be accurate enough for prediction, particularly in problems where capturing characteristics is critical, as in wave propagation problems.  In this paper we propose a modification to sequential state estimation that  can improve 
estimates where features are important. We denote this two-state data assimilation strategy {\it displacement assimilation.}
Many of the sequential data assimilation procedures  yield an estimate of $X(t)$ that may not be in the solution space of the model; for that matter, it might not even be physical. Constraints may be added to the Bayesian statement 
to promote physically reasonable properties in the analysis estimate, which motivates inserting
an extra step into the data assimilation process. 
The method developed here applies this strategy
 to improve the estimates of a variance minimizing strategy when morphological features in the state variable are important. For example,  suppose we want to estimate characteristic paths from a wave process. These paths are both space and time dependent (phase dependent). If designed properly, an assimilation method that makes a phase correction in addition to an estimation might deliver better estimates of such things, as the space-time information of what generated a wave, or the space-time information that better tracks the wave characteristics. This two-stage process is what we call {\it displacement assimilation}.

If the displacement correction and the estimation process are kept separate, it is possible to develop a displacement correction scheme that could then be applied to a variety of different estimation strategies. The development we present here is in that spirit. In this study we will be contrasting the standard enKF and something we call displacement enKF, wherein the difference is that in the latter we add a phase correction.

Adding phase corrections within the context of data assimilation is not a new idea. Among other works, we can mention
\cite{Hoff95}, who argued how this procedure might improve predictions in meteorology (see also  \cite{Brew99}, \cite{Brew02}).   Ravella and collaborators  \cite{Rave07} applied  spatio-temporal nudging to make better predictions of hurricanes.  Percival used ideas from control theory  to find coordinate transformations that could improve predictions \cite{Perc08}; his approach is the closest to our work. 

Percival \cite{Perc08} used area preserving flows to 
suggest the appropriate {\em continuously varying} phase correction, and also defined a thresholding condition to terminate the procedure once sufficient position correction had been performed. In contrast, we will use area preserving maps to do the position correction ``in one shot" every time new data is assimilated into the model. 

One additional criterion in our methodology is to ensure that the phase corrections are kinematically consistent with the underlying physics of the system. This is not always necessary for making phase corrections. In contrast, Frazin
\cite{Fraz12}, for example, makes phase corrections that have no physical basis -- the aim of his work is to improve the  optical data using a data assimilation system. In our context however, it is critical that the corrections be consistent with the underlying physics.

\section{Displacement Assimilation in  Ensemble Vortex Tracking}

The net effect of large-scale forcing on a system of interacting coherent vortices is two-fold. The dominant effect is in modifying the trajectories of (the centers) of the vortices. A much smaller effect is changing the shape/structure of individual vortices \cite{pt-vortex-models}. This naturally suggests the use of statistically derived filtering algorithms to track the trajectories of the vortices for data assimilation.
The trajectories of interacting vortices are highly sensitive to changes in their relative displacement and amplitude, and thus present a useful ``toy model" for developing and testing our displacement assimilation methodology. We will use  a stochastic barotropic vorticity equation to compare the  enKF estimates with and without displacement adjustments. 
Our model is noisy; we do not use the ideal vorticity equation. The added noise accounts for  unresolved or ignored physical processes in the actual 
physical system. Furthermore, there can also be 
 uncertainties in  initial and boundary conditions. The  \emph{stochastic vorticity model} solutions are therefore realizations of the stochastic process, drawn from  a 
 probability distribution. 

Consider an ideal incompressible fluid in a two-dimensional domain $D\subset \mathbb{R}^2$. Denote the velocity vector field $V = (u,v)$, 
  $\nabla \cdot V = 0$. The  vorticity is $\omega = \nabla\times V$. 
 Let  the  stream function be  $\psi$.  $V = \left( -\frac{\partial \psi}{\partial y}, \frac{\partial \psi}{\partial x} \right)$. 
The stochastic model for the vorticity, is
\begin{eqnarray}
 \partial_t\,\omega + V\cdot\nabla\omega &=& \delta f, \quad   \omega:D\times [0,\infty)\to \mathbb{R},  \label{eqn:gen_vort_cons} \\
\omega(z,0) &=& \omega_0(z) + \delta\omega_0(z), \nonumber \\
\nabla^2 \psi &=& \omega, \quad  \psi:D\to \mathbb{R}, \label{eqn:poisson_stream} \\
 \nabla \psi \cdot \partial D &=& \delta b, \nonumber 
\end{eqnarray}
where $\delta f$, $\delta\omega_0$, and $\delta b$ are perturbations in the forcing, initial condition, and boundary condition with known probability distributions. Perturbations in initial condition and boundary conditions will not be invoked in this study. 
The value of the stream function, along the boundary of the domain, $\partial D$
is prescribed.   Note that 
(\ref{eqn:poisson_stream}) couples the the vorticity and stream functions, and consequently couples a set of vorticity features to their Lagrangian paths. The Poisson equation (\ref{eqn:poisson_stream}) is linear, so it maps perturbations in vorticity features directly into perturbations in their trajectory. Accumulated noise in nearby features mutually affects the trajectories of the vortices. Conversely, in systems constrained by diagnostic relationships such as  (\ref{eqn:poisson_stream}), studying spatial displacements in features provides a way to correct the accumulated errors in their amplitude, which improves the subsequent prediction of their future path. 
  Hence, the importance of the forcing error term $\delta f$ which can generate random local changes in the intensity of features, as well as nudge the positions of nearby features.

\begin{figure} \centering
\includegraphics[width=\hsize]{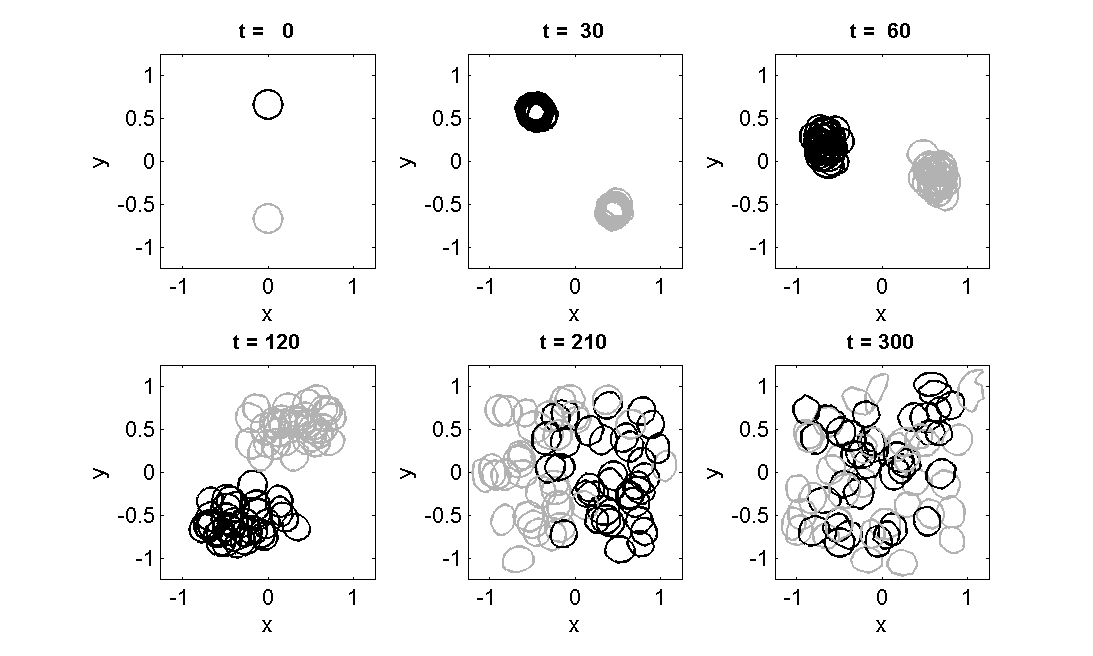}
\caption{\textbf{The noise regime of the model.} A sample of 35 realizations of the vorticity model, with equal-height ($\omega = .5$) contours displayed. The position of the contours becomes homogenized throughout the domain, and the contours are regular and have little variation in area. This shows the Gaussian random process driving the vorticity model causes substantial variation in vortex position and relatively little amplitude error. Moreover, the feature shape is maintained even though the noise breaks the area preservation, a feature of the deterministic vorticity model. }
\label{fig:vortex_homog}
\end{figure}

 \ref{tab:model_params} gives details on the parameter values for the 
 calculations that follow. They were chosen to provide a system of spatially continuous features which experience significant position errors while retaining enough structure to persist throughout the simulation. The initial condition defines a system of co-rotating vortices. In Figure \ref{fig:vortex_homog} two known ($\delta \omega_0 = 0$) and initially identical vorticity pulses are mutually advected in a bounded rectangular domain. The boundary conditions  enforce a zero-outflow condition ($\delta b = 0$) that serves two purposes. The first is to ensure features cannot leave the domain, so that the complexity of the tracking problem remains consistent over time and between independent realizations of the model. The second is that it ensures the energy from the forcing errors cannot leave the system, and increases the variance in vorticity over time. The forcing $\delta f$ 
 is driven by independent Gaussian random functions in the velocity variables with spatial decorrelation on the scale of the vortex features ($\approx 0.25$). This results in low frequency perturbations that modulate the amplitude of each vortex while preserving its geometry so that it can remain coherent throughout the simulation. The amplitude perturbations in the vortices disrupt the stream function governing their oscillation and alter their spatial trajectories, which 
 introduces position errors. Hence the forcing noise places the system in a regime where position variance is the dominant form of model error, and the noise correlation is on a scale that  maintains the feature geometry while allowing each vortex to have independent amplitude perturbations. 

The performance of the standard and displacement-corrected enKF will be compared.  In particular, the decay of forecast and analysis bias with ensemble size is confirmed and compared between both filters. 
A reduction in forecast and analysis variance with respect to the standard method is observed. Moreover, the displacement filter 
 preserves the vortex structure evident in the true vortex system, while the standard enKF introduces spurious deformations. We present an argument explaining this improvement by relating it to the correction of forecast statistics that is a result of the preconditioned filter.

\subsection{Model Setup and Observations}

In the stochastic vorticity dynamics  we assume a (deterministic) initial vorticity condition $\omega(z,0) = \omega_0(z)$ for $z\in D = [x_L,x_U]\times [y_L,y_U]$. The vorticity trajectory $\omega: D\times [0,\infty)\to \mathbb{R}$ is given by
\begin{equation} \label{eqn:vort_prog_model}
\frac{\partial \omega}{\partial t} = -V\cdot \nabla\omega + B_\omega(z)\cdot \xi
\end{equation}
 Equation (\ref{eqn:vort_prog_model}) is subject to zero-outflow boundary condition $\nabla\Psi\cdot \partial D = 0$, where $\Psi: D\to \mathbb{R}$. The model error forcing term $\delta f$ is defined by a Gaussian random function with spatial covariance factor $B_\omega(z)$, and white noise process $\xi\sim {\cal N}(0,I_r)$. Equation (\ref{eqn:vort_prog_model}) is approximated on a uniform rectangular grid on $D$ with $N_x$ and $N_y$ subintervals in the $\{x,u\}$ and $\{y,v\}$ dimensions, respectively, and the vorticity, velocity, and stream function grids are aligned.  Then the total dimensionality of the problem is $N = (N_x+1)(N_y+1)$. This stochastic differential equation (SDE) is integrated in time using the stochastic Heun method \cite{KP}. Each time the right-hand side of Equation (\ref{eqn:vort_prog_model}) is evaluated, the stream function is obtained by solving the Poisson equation $\Delta \Psi = \omega$ via a second-order central difference discretization of the Laplacian, and the velocity components are computed using second-order centered difference approximations of the derivatives.

The structure of $B_\omega$ and dimension $r$ of the Gaussian random function driving the model noise are inferred from assumptions on the variability of the velocity components. Since vorticity is a derived state parameter, we make an effort to define a (reasonable) model uncertainty based only on the (given) covariance statistics in the observable state variables, i.e. the velocity vector field components. The velocity components are assumed mutually independent and spatially distributed with a squared-exponential correlation structure,
\begin{equation}
k_u(z_1,z_2) = k_v(z_1,z_2) = \sigma_V^2 \exp\left( -r_V^{-2}\left\| z_2 - z_1 \right\|_2^2 \right)
\end{equation}
where $\sigma_V^2$ and $r_V$ are the scale and shape parameters setting the pointwise variance and decorrelation length of the model error kernels $k_u$ and $k_v$. Let $Q_V$ denote the Grammian matrix corresponding to the evaluation of the kernels on the discretized domain $D$, and compute its factorization $Q_V = B_V B_V^T$, where $\mbox{rank}(B_V)$ is sufficient to represent all numerically significant modes of the original covariance matrix. Each spatial mode of each velocity component is then dampened to zero at the boundaries in the outflow directions, so that the model perturbations are guaranteed consistent with the boundary conditions. Specifically, another shape parameter $r_b$ is defined to scale the width of the transition region. Then the following bump function is defined,
\begin{equation}
\left[ 1 - \exp\left(-\frac{|x-x_L|}{r_b}\right) \right]\cdot \left[ 1 - \exp\left(-\frac{|x-x_U|}{r_b}\right) \right]
\end{equation}
which is unity in $[x_L+r_b,x_U-r_b]$, and scales the $u$ model noise modes. A similar bump function in the $y$ dimension is defined and applied to the $v$ modes. With consistent velocity perturbations, the vorticity perturbation modes can be computed from the definition $\omega = \partial_x v - \partial_y u$,
\begin{equation}
\delta\omega = B_\omega\cdot \xi = \left[ \begin{array}{cc} \frac{\partial B_v}{\partial x} & -\frac{\partial B_u}{\partial y} \end{array} \right] \cdot 
\left( \begin{array}{cc} \xi_u & \xi_v \end{array} \right)
\end{equation}
where the derivatives are approximated by centered differences. The vorticity model perturbations driving the SDE in Equation (\ref{eqn:vort_prog_model}) follow a Gaussian random function with spatial covariance $Q_\omega = B_\omega B_\omega^T$. We should keep in mind that this covariance actually measures noise from velocity perturbations in a $r = 2\cdot \mbox{rank}(B_V)$-dimensional (observation) space expressed in $r/2$-dimensional vorticity model space.

In our simulations, the initial condition was deterministic and  consisted of two co-rotating vortices,
\begin{equation}
\omega_0(z) = a_1\cos^2\left[ \frac{\pi}{2} r_1(z) \right] + a_2\cos^2\left[ \frac{\pi}{2} r_2(z) \right]
\end{equation}
where the rescaled radius $r_j(z)$ defines the support of each vortex, and is given by
\begin{equation}
r_j(z) = \frac{d_j(z)}{r_{s,j}} \chi_{\left\{ d_j(z) \leq r_{s,j} \right\}}(z)
\end{equation}
and $d_j(z) = \left[ \left(x - x_{c,j}\right)^2 + \left(y - y_{c,j}\right)^2 \right]^{1/2}$, the squared distance from the vortex, center, and $\chi_A(z)$ is the indicator function for all $z\in A$. The trajectory of one realization of the generalized model, including the initial condition, is provided in Figure \ref{fig:ex_truth_obs}, with one frame for each assimilation time.

The observations of the state variable in the present example will be  a  sparse array of noisy Eulerian observations of the velocity.  Let $\{(x_{d,j},y_{d,j})\}$ for $1\leq j\leq N_d$ denote the set of observed locations. We will take them to be an approximately uniform $(N_{d,x}+1)\times (N_{d,y}+1)$ array in $D$, though for posing the problem they can be  considered scattered. For convenience, they will be a  subset of the already discretized coordinates on which the model is defined.  Let the true state of the vorticity field be denoted by $\omega^t$, and define the linear velocity amplitude observation operator to be
\begin{equation}
h_a(\omega;\{x_{d,j},y_{d,j}\}) = \left[ \begin{array}{c} -\partial_y \Delta^{-1}\omega\left(\{x_{d,j},y_{d,j}\}\right) \\ \partial_x \Delta^{-1}\omega\left(\{x_{d,j},y_{d,j}\}\right) \end{array} \right]
\end{equation}
Then the velocity observations are given by $d_V := (d_u,d_v)^T = h_a(\omega^t) + \epsilon_d$, where $\epsilon_d\sim {\cal N}(0,R)$ are the assumed independent and identically distributed normal observation errors, so that $R = \tau I_{2N_d}$.

\subsection{Displacement Correction Via Area-Preserving Maps}
The stochastic vorticity equation
 produces a distribution of vorticity functions whose features have perturbed amplitudes; this introduces noise in the Lagrangian path of these features. 
The generalized inverse problem for position errors is to identify which of the possible Lagrangian perturbations would produce a realization of the stochastic vorticity function that most closely matches the true state of the system. Consider a quadratic penalty functional that measures the Eulerian distance between the true vorticity function $\omega^t$ and various candidate analyses $\omega$ with respect to a specific norm $\|\cdot \|$,
\begin{equation}
 \label{eqn:map_functional}
\mathcal{J}_a[\omega] = \frac{1}{2}\left\| \omega^t(Z) - \omega(Z) \right\|^2
\end{equation}
In order to relate the forecast prediction to candidate analyses through corrections in the position of features, we define a family of smooth, invertible maps $Z = \Phi(z;a)$. Throughout the following discussion we will consistently refer to the coordinates native to the true and observed vorticity function by $Z$, and the coordinates on which the forecast model $\omega^f$ is defined by $z$. Then we can write $\omega(Z) = \omega^f(z) = \omega^f\left[ \Phi^{-1}(Z;a) \right]$. Now we can recast  (\ref{eqn:map_functional}) in terms of the map parameters,
\begin{equation}
\mathcal{J}_p[a] = \frac{1}{2}\left\| \omega^t(Z) - \omega^f\circ\Phi^{-1}(Z;a) \right\|^2,
 \label{eqn:map_param_functional}
\end{equation}
which allows us to ask which position corrections provide the best agreement (minimum $\mathcal{J}_p$) with measurements of the true vorticity function. 
 If the maps $\Phi$ are to model stream function perturbations, they also must be divergence-free, in the sense that they preserve the total vorticity on any subset as it is mapped from one position to another. These subsets include any closed contour of the vorticity function, including the concentric and parallel contours that describe the geometry of features. Hence, preserving the total vorticity and area on contours is akin to preserving the geometry of features, i.e quantities like volume and intensity. 

\subsubsection{Symmetric Parameterization of Displacement Maps}

Canonical transformations are perturbations of the identity map, and they can be used to derive a parameterization which utilizes more familiar tools. Let $\zeta(X,y;a)$, which will be referred to as the \emph{map function}, be an at least twice continuously differentiable function, where $a$ are constant parameters to be defined later. Let $G_0(X,y) = -Xy + \zeta(X,y;a)$, then the transformation equations can be written as.
\begin{eqnarray} \label{eqn:canonical_trans}
x = X + \frac{\partial}{\partial y} \zeta(X,y;a)  \quad Y = y + \frac{\partial}{\partial X} \zeta(X,y;a) .
\end{eqnarray}
It can be verified that the map defined by Equations (\ref{eqn:canonical_trans}) is indeed is area preserving by computing its gradient $F = \nabla_{(x,y)}(X,Y)$ and showing $|F| = |F^{-1}| = 1$. In continuum mechanics, $F$ is known as the \emph{deformation gradient}, and is just the Jacobian matrix of the coordinate transformation. It is useful in computing strains induced by the coordinate system, which will be employed to regularize the estimation of an optimal displacement map in a later section. In particular, $\zeta$ will be modeled as a Gaussian random function, $\zeta(X,y,a) = \sum_j^N a_j B_j(X,y)$, for a suitable function basis $\{B_j\}$, and $\left(a_1, ..., a_N\right)$ is drawn from a multivariate normal distribution.

If we solve for the displacements $X-x$ and $Y-y$ in Equation (\ref{eqn:canonical_trans}), we recover a system which approximately defines $\zeta$ to be a stream function for a velocity field parallel to the spatial increments. In the limit of small displacements, canonical transformations are equivalent to a particular divergence-free flow along stream function contours. The transformation equations can be rewritten so that the map $\Phi$ is an approximate cross-section of the flow. For a small time parameter $0 < \epsilon \sim \|\nabla \zeta\| \ll 1$, define $\zeta_\epsilon(X,y) = \epsilon \Psi(x,y)$. Then,
\begin{eqnarray}
\frac{\delta x}{\delta t} &=& \frac{X-x}{\epsilon} = -\frac{1}{\epsilon}\frac{\partial \zeta_\epsilon}{\partial y}(X,y)  \nonumber  \\
\frac{\delta y}{\delta t} &=& \frac{Y-y}{\epsilon} = \frac{1}{\epsilon}\frac{\partial \zeta_\epsilon}{\partial X}(X,y)  \nonumber
\end{eqnarray}
As  $\epsilon \rightarrow 0$ (a formal limit) we get
\begin{eqnarray}
\frac{\delta x}{\delta t} &=&  -\frac{\partial}{\partial y}\Psi(x,y) \nonumber \\
\frac{\delta y}{\delta t} &=& \frac{\partial}{\partial x} \Psi(x,y).
\end{eqnarray}
and in the limit of small displacements, $\epsilon\to 0$, $z\to Z$, and the system converges to the stream function definition for a conservative velocity field. This motivates us to define area preserving maps $(X,Y) = \Phi(x,y;a)$ by integrating stream functions over a fixed time interval,
\begin{eqnarray}
 \label{eqn:stream_func_map}
&& \Phi(x,y;a) = (x,y) + \nonumber \\
&&\int_0^1 \left( -\Psi_y\left[ x(t), y(t); a \right] , \Psi_x\left[ x(t), y(t); a \right] \right) dt
\end{eqnarray}
The parameterization of $\Psi$ is linear in $a$, i.e. taking $\Psi(x,y;a) = \sum_j^N a_j B_j(x,y)$, then the inverse map $(x,y) = \Phi^{-1}(X,Y;a)$ is obtained simply by making the substitutions $(x,y,a) \to (X,Y,-a)$. Unlike the usual formulations of canonical transformations, the approach of integrating a flow gives an explicit, rather than an implicit function, and is un-constrained, unlike the constraints on the mixed partial derivatives of a generating function to guarantee the invertibility of  a canonical transformation be invertible \cite{SV-KAM}. For a given position correction problem, a sequence of maps can be defined through iterative optimization of functionals like Equation (\ref{eqn:map_param_functional}). In this context, the small displacement assumption is reasonable, and we can proceed to take advantage of the theoretical simplicity of the canonical map structure to develop additional tools. Moreover, to generate the maps we take advantage of  the practical convenience of integrating divergence-free flows. Hereafter we identify $\Psi(x,y;a) \leftrightarrow \zeta(X,y;a)$, and the particular usage will be clear from context. In fact, these two approaches are identical in applications which require the map to be linearized, such as when we project the statistics of amplitude perturbations in vorticity onto the space of displacement map parameters. In the next section we apply both points of view in order to derive regularization terms that help provide smooth displacements where feature information is present, based on an application of continuum mechanics.

\subsubsection{Transformation Regularization}
In the set of possible feature displacement maps, we want to avoid coordinate transformations which excessively deform the geometry of strong features in the model function. In fact, excessive deformations in general should be avoided. Since the likelihood functional will penalize only coordinate transformations where there is significant feature information, then an otherwise unfettered optimization routine will dissect regions where there are no features in either the observed or modeled functions. Perturbations in lightly-constrained regions can generate enough of a change in the penalty functional to create local depressions in the objective function that have nothing to do with the displacement of significant features--precisely the non-convexity that  regularization schemes, such as Tikhonov regularization, are designed to eliminate. 

 Consider the  problem of determining the state $y$ which best explains a set of noisy observations $d = h(y^t) + \epsilon$ of an unknown true state $y^t$, where the nonlinear observation operator can be modeled by $h(y)$ and the observation noise $\epsilon\sim {\cal N}(0,R)$. We want to solve the ill-posed problem $d = h(y)$, and resort to minimizing the likelihood functional,
\begin{equation}
\mathcal{J}_L[y] = \left\| d - h(y) \right\|_{R^{-1}}^2
\end{equation}
where the so-called maximum likelihood estimator is the maximizer of $\exp^{-\mathcal{J}_L[y]}$. Next, presume we have prior information that $y^t\sim {\cal N}(y^f,P^f)$, where $y^f$ is a forecast mean, and $P^f$ the forecast covariance. The maximum-a-posteriori estimator is the maximizer of $\exp^{-\mathcal{J}_P[y]}$, where
\begin{equation} \label{eqn:post_fcnl}
\mathcal{J}_P[y] = \mathcal{J}_L[y] + \left\| y - y^f \right\|_{(P^f)^{-1}}^2
\end{equation}
We are penalizing the distance from the forecast $y^f$ in the matrix-weighted norm defined by the inverse of the forecast covariance, $(P^f)^{-1}$. The covariance matrices in real applications are often singular, by design or to machine precision. The inverse need not exist, let alone have a diagonal 
factorization.

 Suppose the displacement map complexity is kept well below that of the original problem,
\begin{equation} \label{eqn:dim_cond}
\mbox{dim}(a)\ll \mbox{dim}(\omega) = \mbox{dim}(\zeta) = \mbox{dim}(\Phi)
\end{equation}
Denote the linear map from the displacement parameters to vorticity values as $T_{\omega a}$, and denote its pseudoinverse $T_{a \omega}$, the map projecting a vorticity function into displacement parameter space. These maps can be used to estimate a prior model covariance of map parameters from a given prior vorticity covariance. For vorticity perturbations $\delta \omega$ and map parameter perturbations $\delta a$, we can write
\begin{eqnarray}
P_\omega 	&=& \mbox{Cov}(\delta \omega) = \mathbb{E}(\delta\omega \delta\omega^T)  \nonumber \\
&=&  \mathbb{E}(T_{\omega a}\delta a\delta a^T T_{\omega a}^T) \nonumber \\
		&=& T_{\omega a}  \mathbb{E}(\delta a\delta a^T) T_{\omega a}^T  \nonumber \\
		&=& T_{\omega a} \mbox{Cov}(\delta a) T_{\omega a}^T = T_{\omega a} P_a T_{\omega a}^T
\end{eqnarray}
and
\begin{equation} \label{eqn:pos_covar_proj}
P_a = T_{a \omega} T_{\omega a} P_a T_{\omega a}^T T_{a \omega}^T = T_{a \omega} P_\omega T_{a \omega}^T
\end{equation}
Then we can consider regularization terms defined directly in the space of displacement parameters. In particular, we write
\begin{equation} \label{eqn:reg_fcnl}
J_p[a] = \frac{1}{2} \left\| d_\omega - \omega^f\circ \Phi^{-1}(Z_d,a) \right\|_{R^{-1}}^2 + S[a;\alpha]
\end{equation}
where we could take the regularization term $S[a] = \| a - a^f \|_{C_a^{-1}}^2$, the usual quadratic form for the model prior, with the forecast $a^f = 0$, and the covariance penalty factor $C_a$. Note that in practice the functional will not be expressed in terms of direct vorticity observations, $d_\omega$, but rather with velocity measurements, $d_V$.
 Alternatively, we can choose a regularization term which includes more information than just the weighted distance from the forecast, such as a preference for smoothly varying maps.

Since we can write down the formula for a coordiante transformation $z = \Phi^{-1}(Z)$, then we can write down the local displacement induced by the transformation, $U(Z) = Z - z(Z) = Z - \Phi^{-1}(Z)$. This is the amount the original coordinate system was deformed to arrive at the current coordiantes $Z$. We also can write down how differential length elements in the former system $dz$ are deformed in the new system $dZ$.  For sufficiently smooth displacements, 
and  $|\zeta| \ll 1$,
\begin{equation}
\epsilon := \frac{1}{2}\left[ \nabla_Z U + (\nabla_Z U)^T \right]
\end{equation}
or in terms of the map function, $\zeta$, the displacement gradient and strain tensor simplify to
\begin{eqnarray} 
\epsilon[\zeta] &=& \left[ \begin{array}{cc} \epsilon_{XX}[\zeta] & \epsilon_{XY}[\zeta] \\ \epsilon_{YX}[\zeta] & \epsilon_{YY}[\zeta] \end{array} \right]  \nonumber \\
& \approx & \left[ \begin{array}{cc} -\zeta_{yX} & \frac{1}{2} ( \zeta_{XX}-\zeta_{yy} ) \\ \frac{1}{2} ( \zeta_{XX}-
\zeta_{yy}) & \zeta_{yX} \end{array} \right].
\label{eqn:eng_strain_tensor}
\end{eqnarray}

 For the strain regularization treatment, we augmented (\ref{eqn:map_param_functional})
 with a weighted Frobenius norm of the engineering strain tensor in (\ref{eqn:eng_strain_tensor}) at each model point,
\begin{eqnarray}
\left\| \epsilon[\zeta] \right\|_{F,\alpha}^2 
	&= &\frac{\alpha_n}{2} \left( \epsilon_{XX}^2[\zeta] + \epsilon_{YY}^2[\zeta] \right)  \nonumber \\
	&+&  \frac{\alpha_s}{2} \left( \epsilon_{XY}^2[\zeta] + \alpha_{YX}^2[\zeta] \right) \nonumber \\
	&= &\alpha_n \left\| \zeta_{yX} \right\|^2 + \frac{\alpha_s}{4} \left\| \zeta_{XX}-\zeta_{yy} \right\|^2 \nonumber \\
	&\sim &\alpha_n \left\| \Psi_{yx} \right\|^2 + \frac{\alpha_s}{4} \left\| \Psi_{xx}-\Psi_{yy} \right\|^2
\end{eqnarray}
where $\alpha_n$ and $\alpha_s$ calibrate the size of the normal and shear strain penalties, respectively. Since there is no prior model distribution limiting the displacement, the optimization algorithm is free to find the true bulk position shift. However, without a calibrated regularization term, the algorithm also is free to suggest local deformations wherever there is not significant feature information. This situation can result from not having any background error at all, as is the case here, or whenever observation uncertainty or scarsity effectively places observation uncertainty on the same or greater scale as the background signal noise. 

Provided a linear parameterization of the stream function $\Psi(z) = \sum_j^N a_jB_j(z) = B(z)\cdot a$, the strain regularization term can be written as a matrix-weighted norm,
\begin{eqnarray}
\left\| \epsilon[\Psi] \right\|_{F,\alpha}^2
	&=& \alpha_n \left( B_{yx} \cdot a\right)^T \left( B_{yx}\cdot a \right) \nonumber \\
	&+& \frac{\alpha_s}{4} \left( B_{xx}\cdot a - B_{yy}\cdot a \right)^T \left( B_{xx}\cdot a - B_{yy}\cdot a \right) \nonumber \\
	&=& \left\| a \right\|_{ C_{\epsilon,\alpha}^{-1} }^2 \label{eqn:pos_reg_term}
\end{eqnarray}
where
\begin{equation} \label{eqn:strain_covar}
C_{\epsilon,\alpha} = \left[ \alpha_n B_{yx}^T \cdot B_{yx} + \frac{\alpha_s}{4} \left( B_{xx}-B_{yy} \right)^T \left( B_{xx}-B_{yy} \right) \right]^{-1}
\end{equation}
If this constraint is simply added to the penalty functional in (\ref{eqn:reg_fcnl}), the standard least-squares minimizer requires the sum $C_a^{-1} + C_{\epsilon,\alpha}^{-1}$, forcing us to explicitly calculate the model prior covariance inverse.
Alternatively, we directly regularize the model covariance
\begin{equation} \label{eqn:map_param_reg}
C_{a,\alpha} = C_a + C_{\epsilon,\alpha}
\end{equation}
so that only the deformation gradient information in  (\ref{eqn:strain_covar}) need be inverted. This is no problem provided the complexity condition in  (\ref{eqn:dim_cond}) is maintained, which depends on the manner in which the displacement maps are discretized. 

\begin{figure} \centering
(a)\includegraphics[height=3in,clip=true,trim=40 0 40 0]{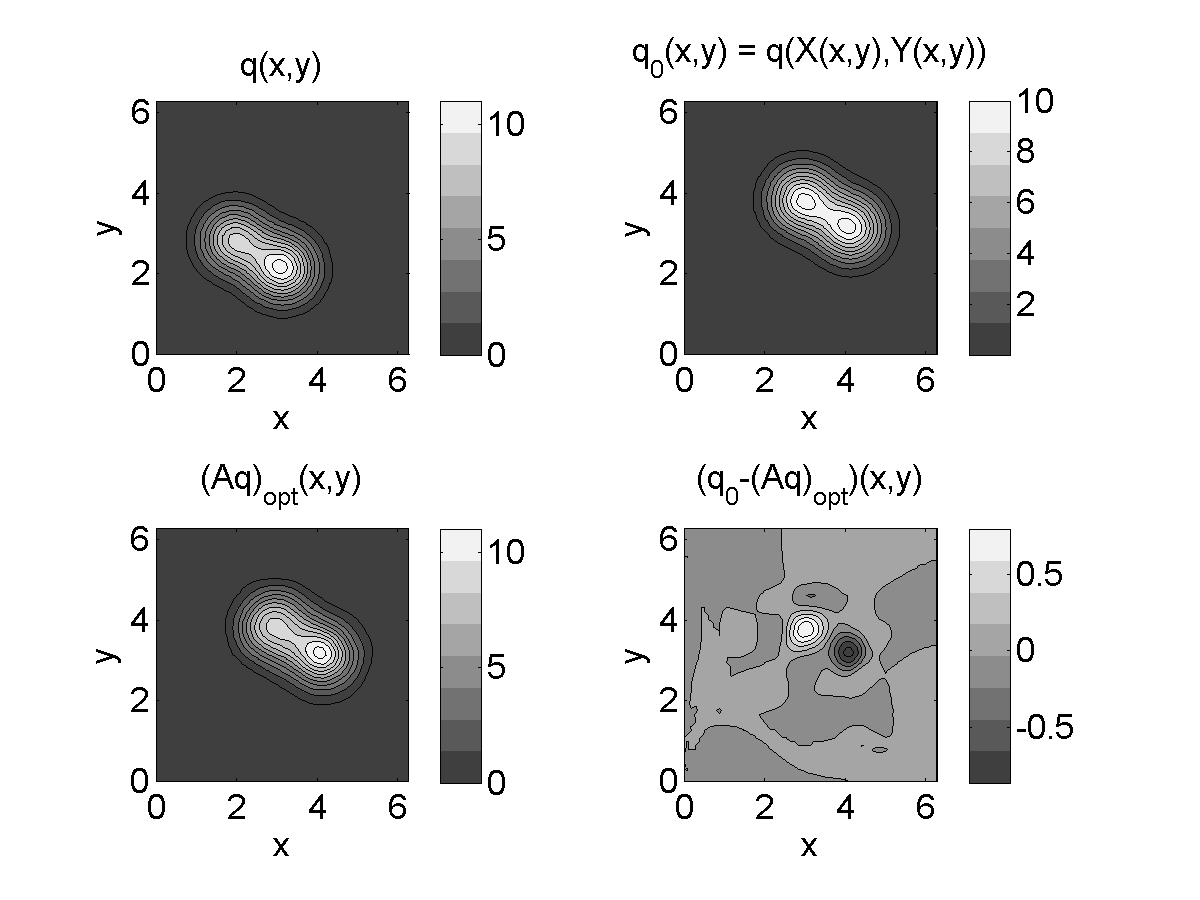} 
(b)\includegraphics[height=3in,clip=true,trim=40 0 40 0]{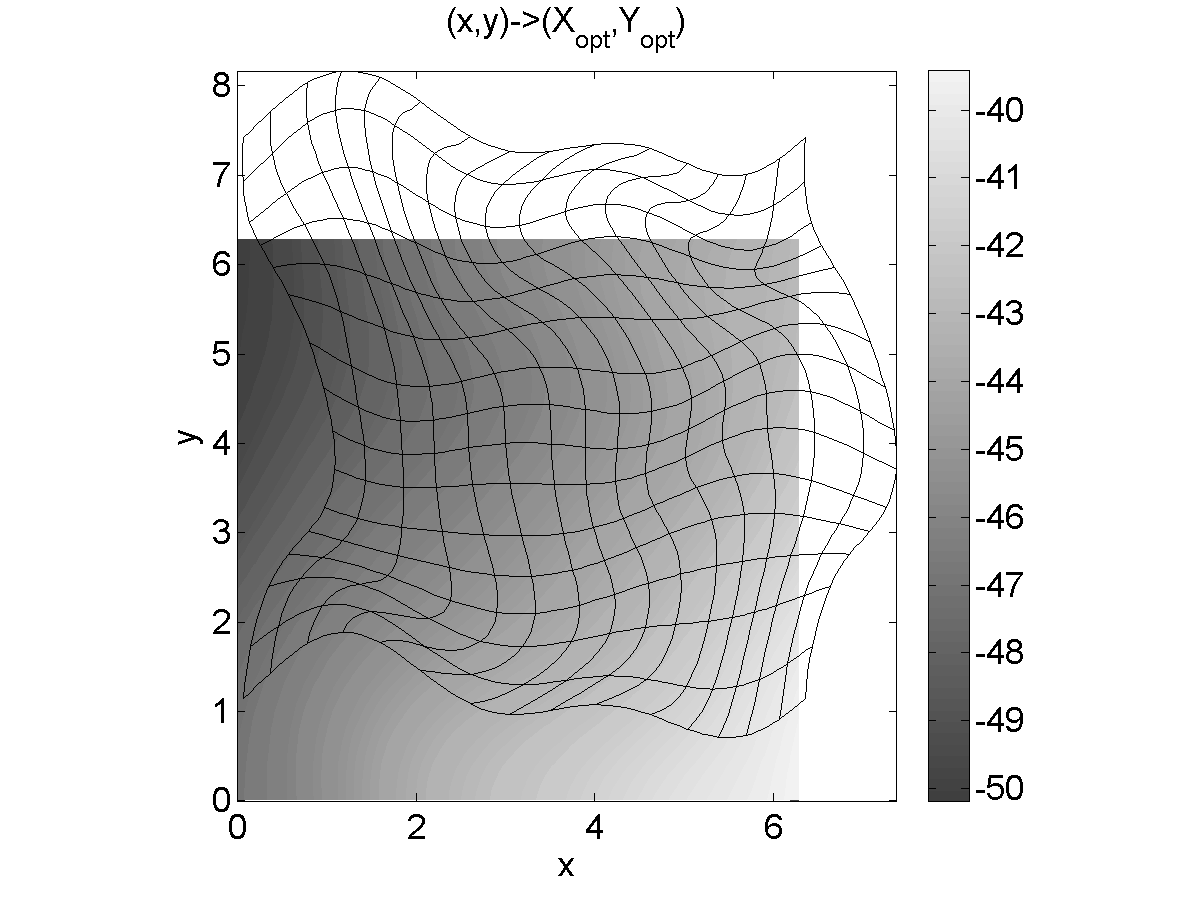} \\
\caption{\textbf{Position realignment} (a) a forecast of a prominent feature with no background noise (upper-left), the true feature with amplitude perturbations (upper-right), an optimal realignment obtained by a standard constrained numerical optimization code (lower-left), and the residual between the optimization and the true feature (lower-right); (b) the displacement maps generated for the realignment as a mesh overlaying the corresponding map function $\zeta$, no regularization.}
 \label{fig:strain_reg_expsa}
\end{figure}

\begin{figure} \centering
(a)\includegraphics[height=3in,clip=true,trim=40 0 40 0]{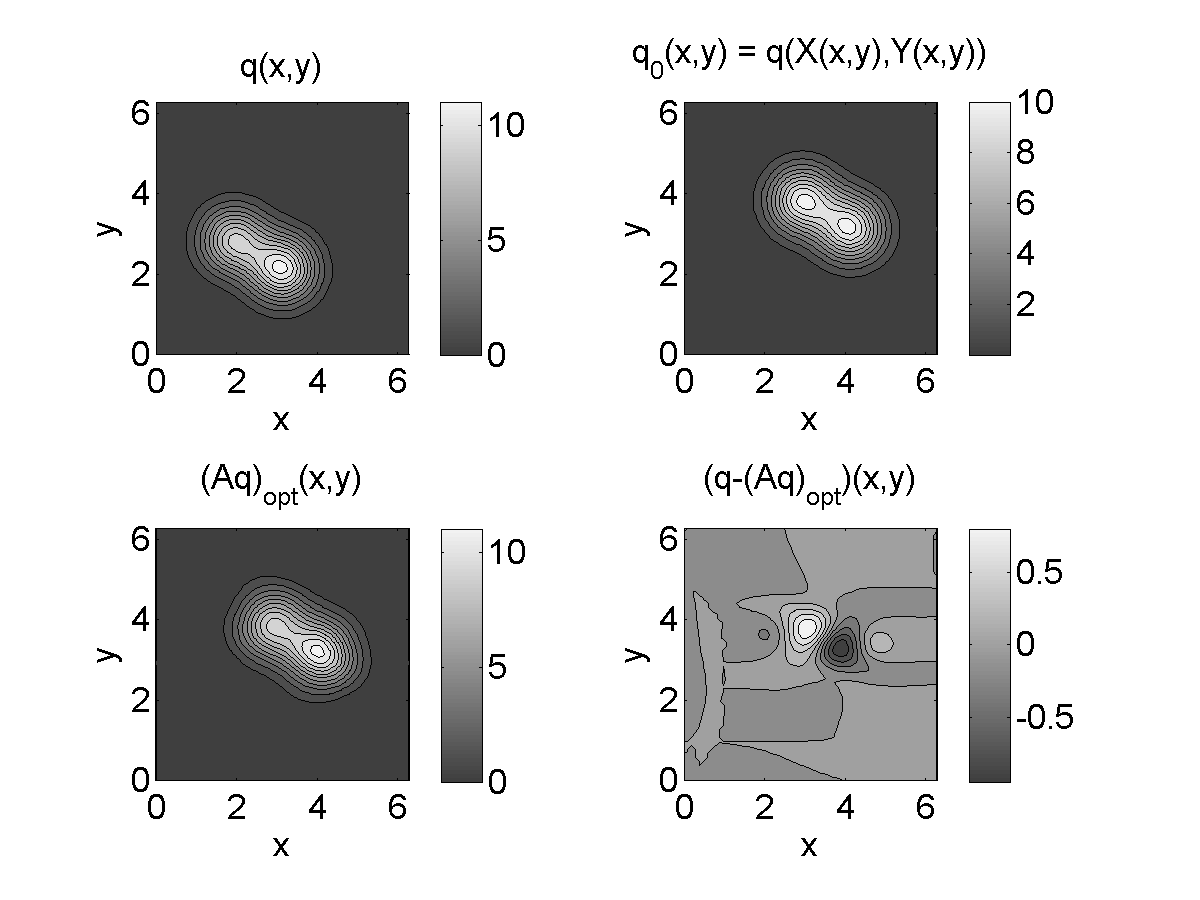}
(b)\includegraphics[height=3in,clip=true,trim=40 0 40 0]{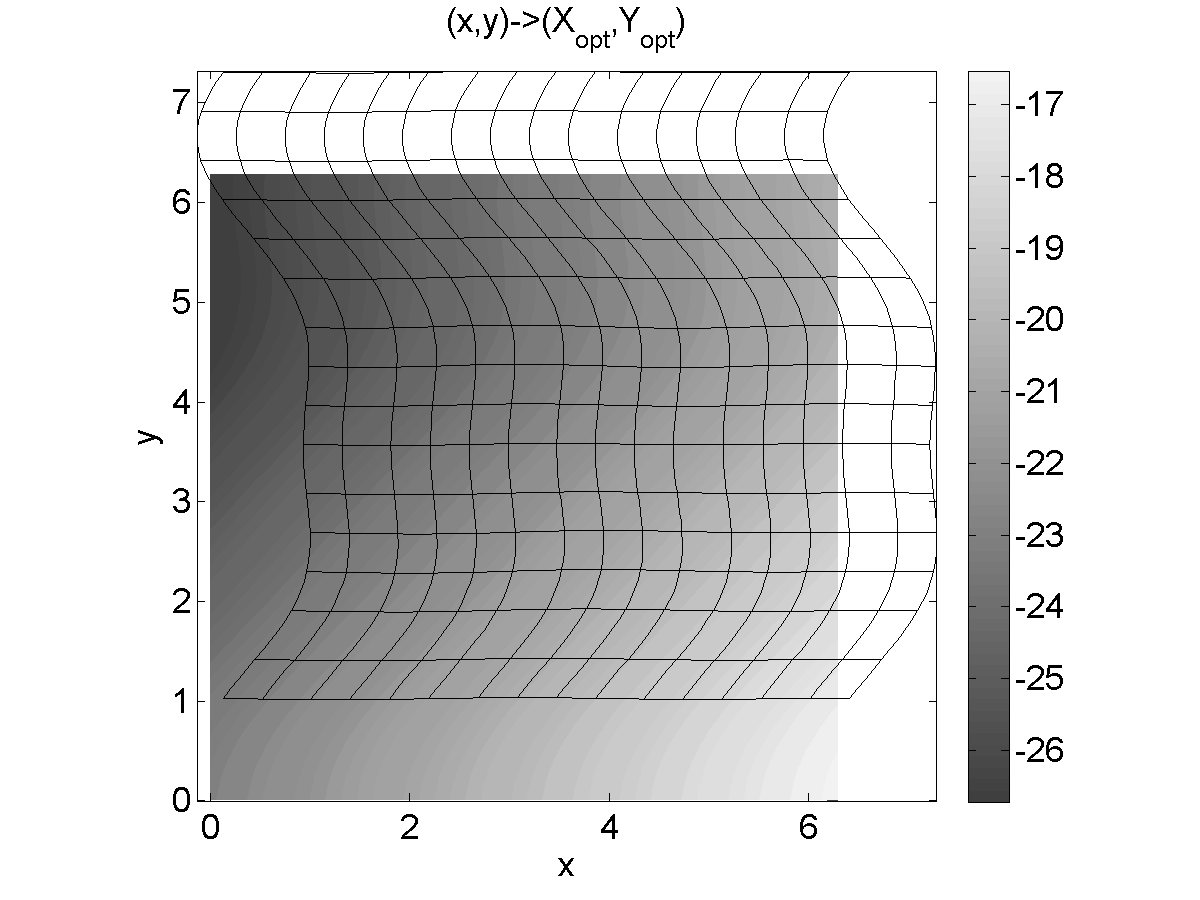}
\caption{\textbf{Position realignment, with strain regularization.} 
Same panels as Figure \ref{fig:strain_reg_expsa}, except that displacement maps have been regularized by the inclusion of a strain regularization term that penalizes spurious deformations in coordinates.
(a) a forecast of a prominent feature with no background noise (upper-left), the true feature with amplitude perturbations (upper-right), an optimal realignment obtained by a standard constrained numerical optimization code (lower-left), and the residual between the optimization and the true feature (lower-right); (b) the displacement maps generated for the realignment as a mesh overlaying the corresponding map function $\zeta$, with regularization.}
 \label{fig:strain_reg_expsb}
\end{figure}

\begin{figure} \centering
\begin{tabular}{rl}
\includegraphics[height = 2in, clip = true, trim = 20 0 30 0]{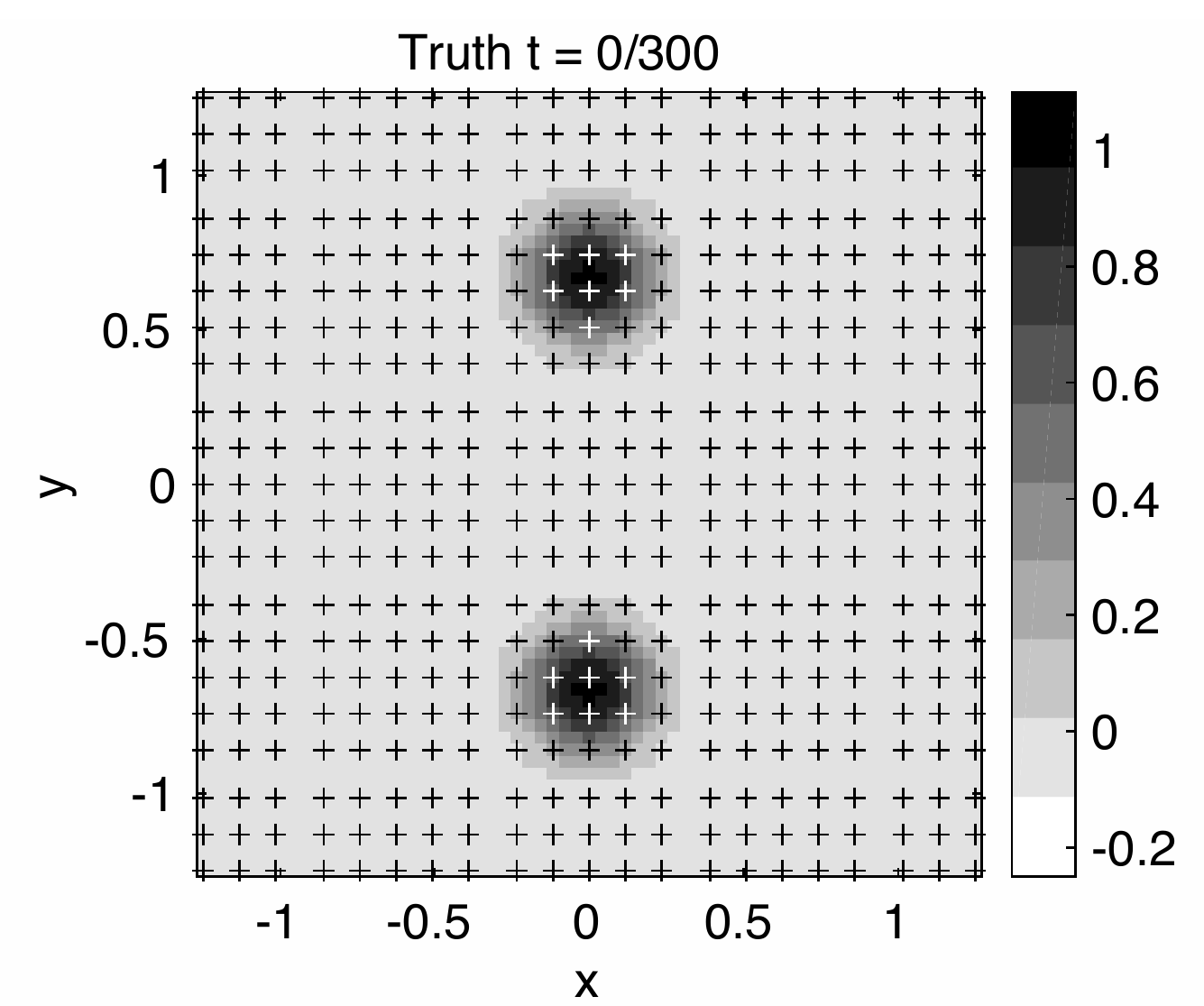} &
\includegraphics[height = 2in, clip = true, trim = 20 0 30 0]{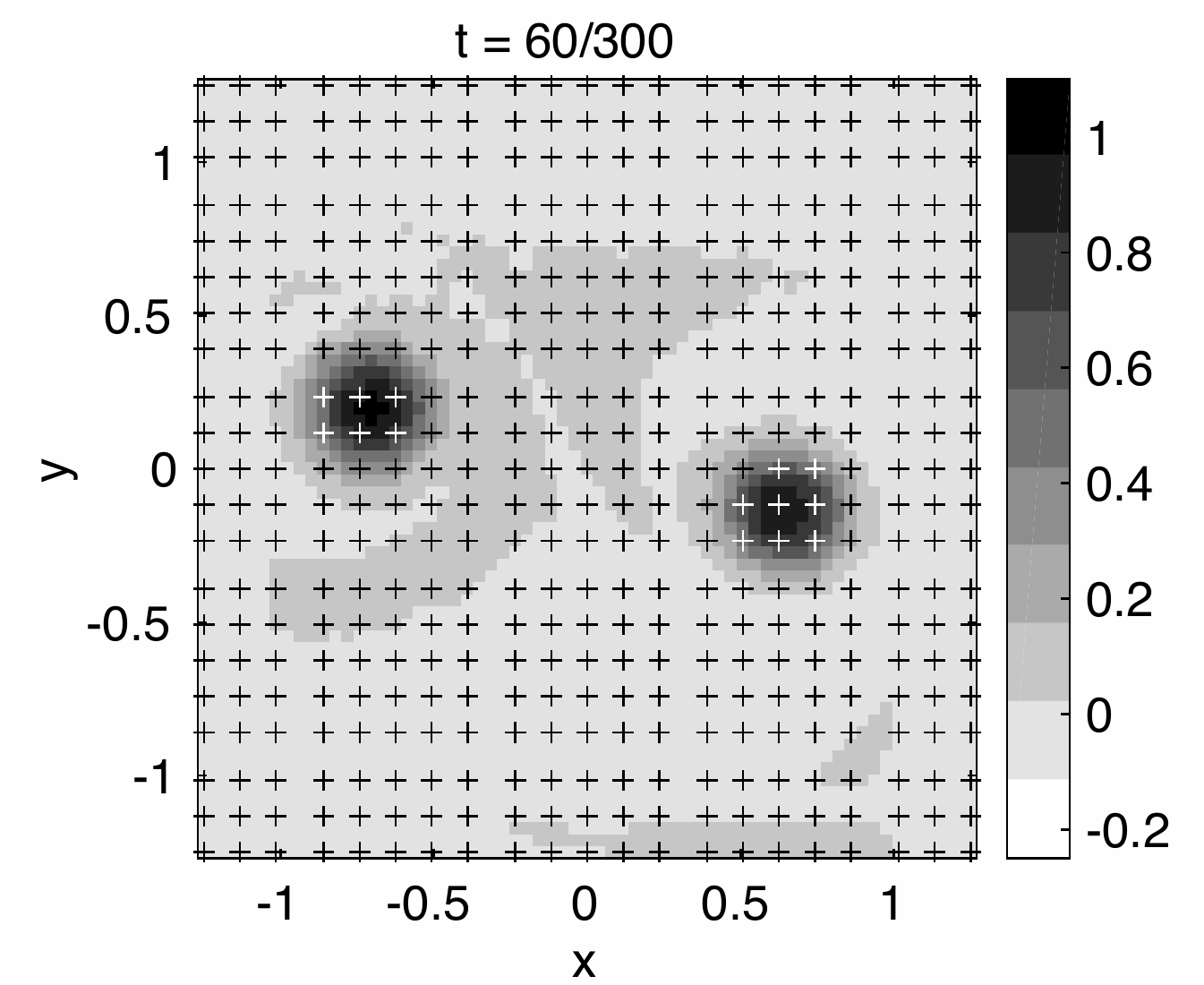} \\
\includegraphics[height = 2in, clip = true, trim = 20 0 30 0]{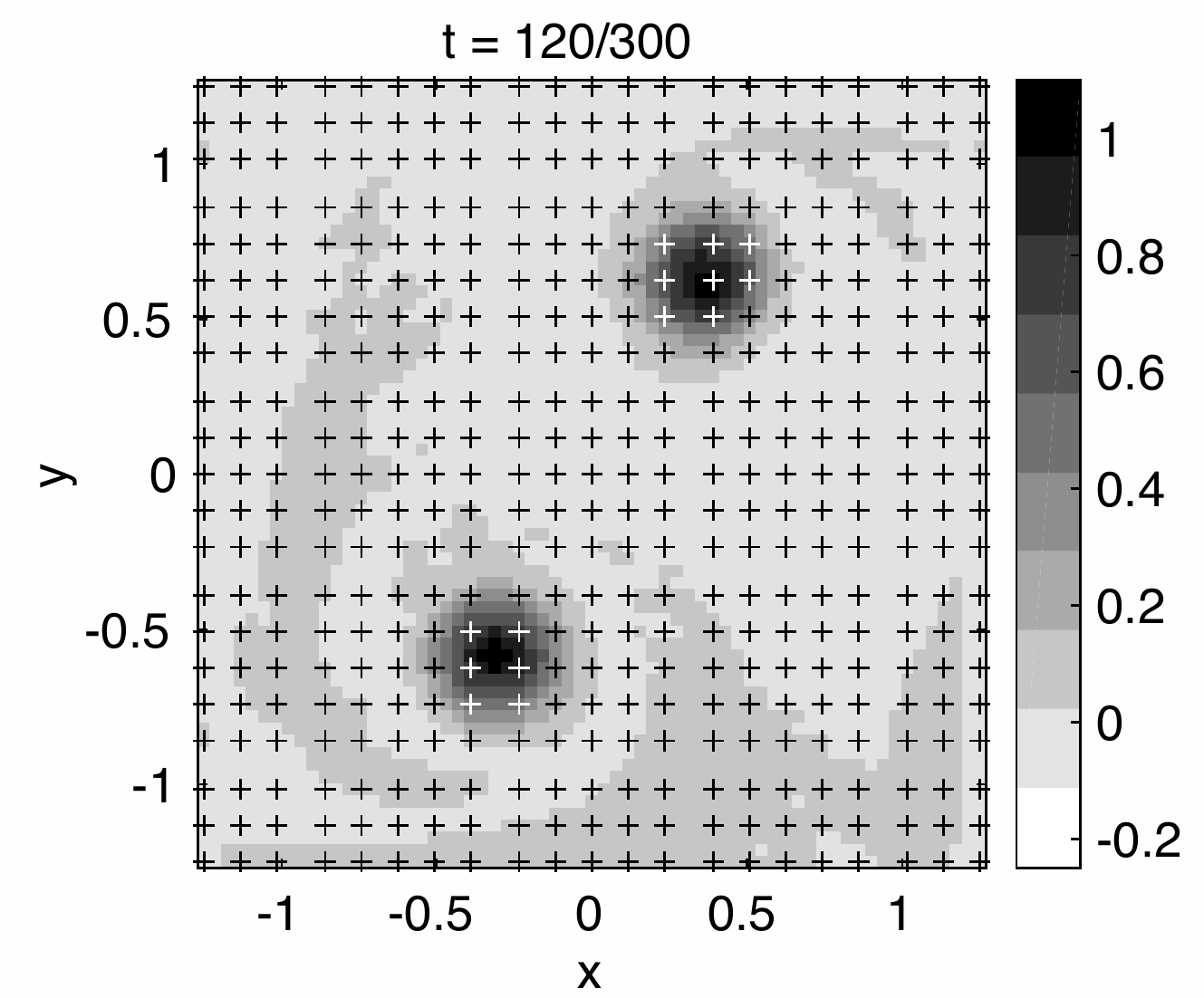} &
\includegraphics[height = 2in, clip = true, trim = 20 0 30 0]{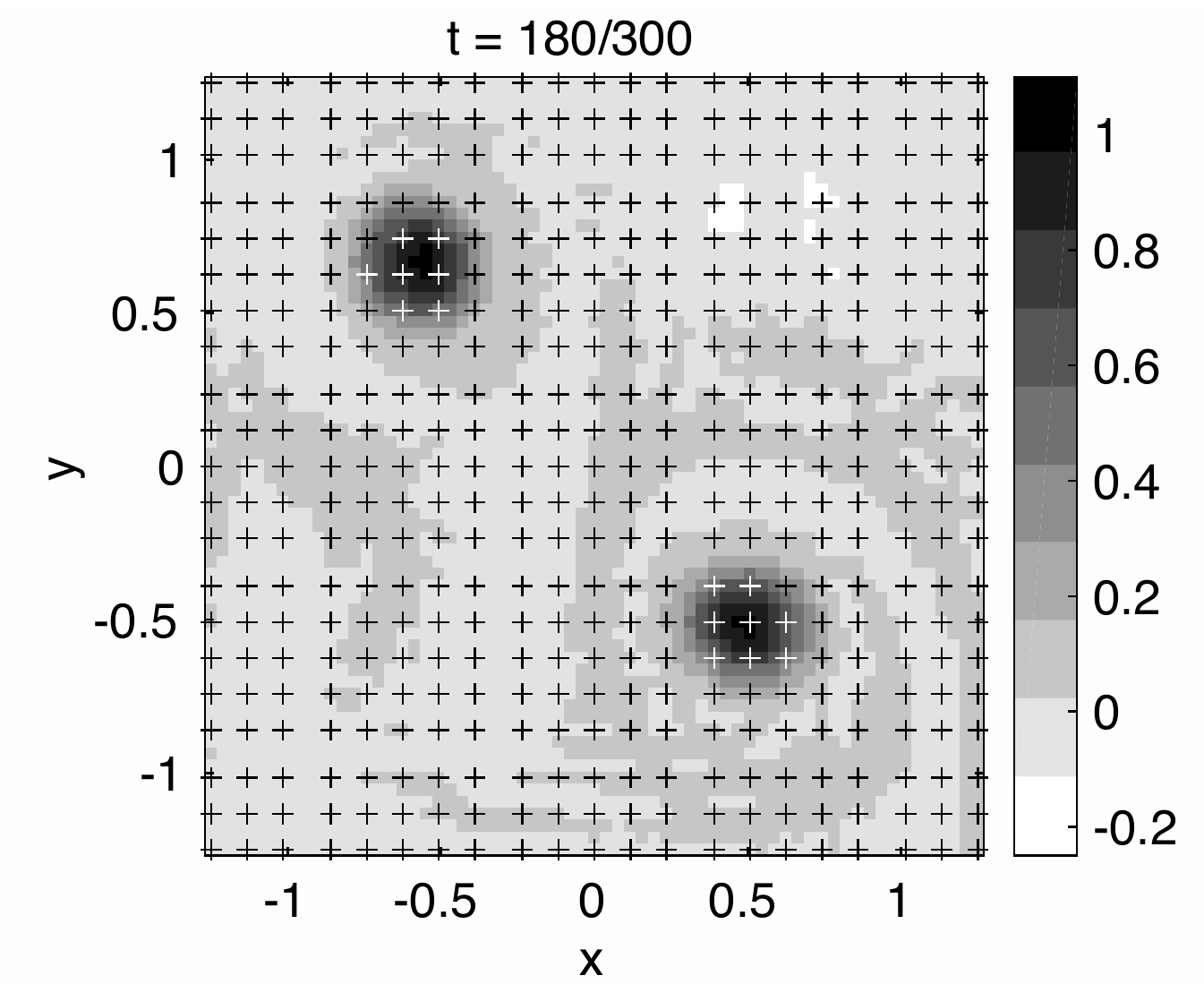} \\
\includegraphics[height = 2in, clip = true, trim = 20 0 30 0]{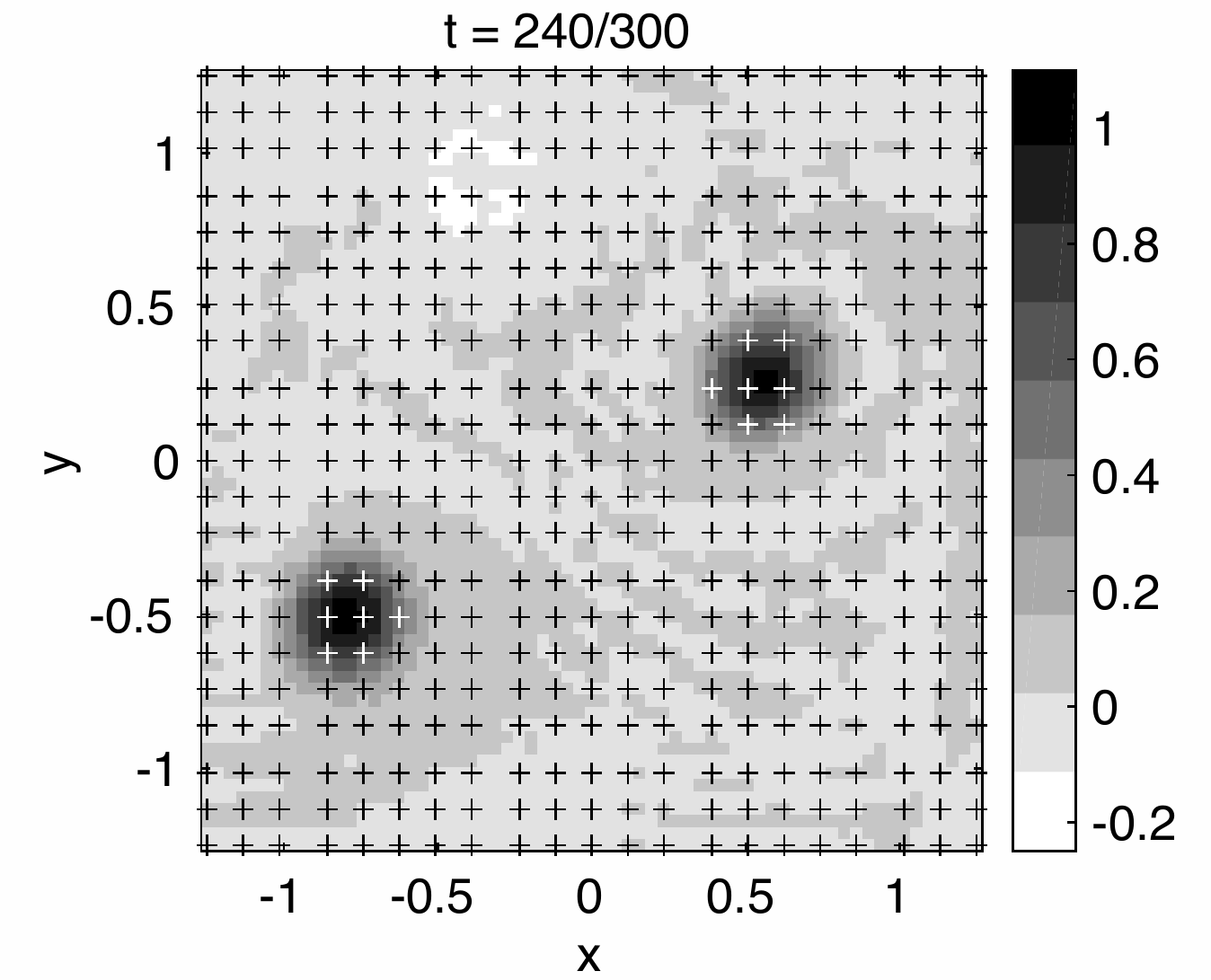} &
\includegraphics[height = 2in, clip = true, trim = 20 0 30 0]{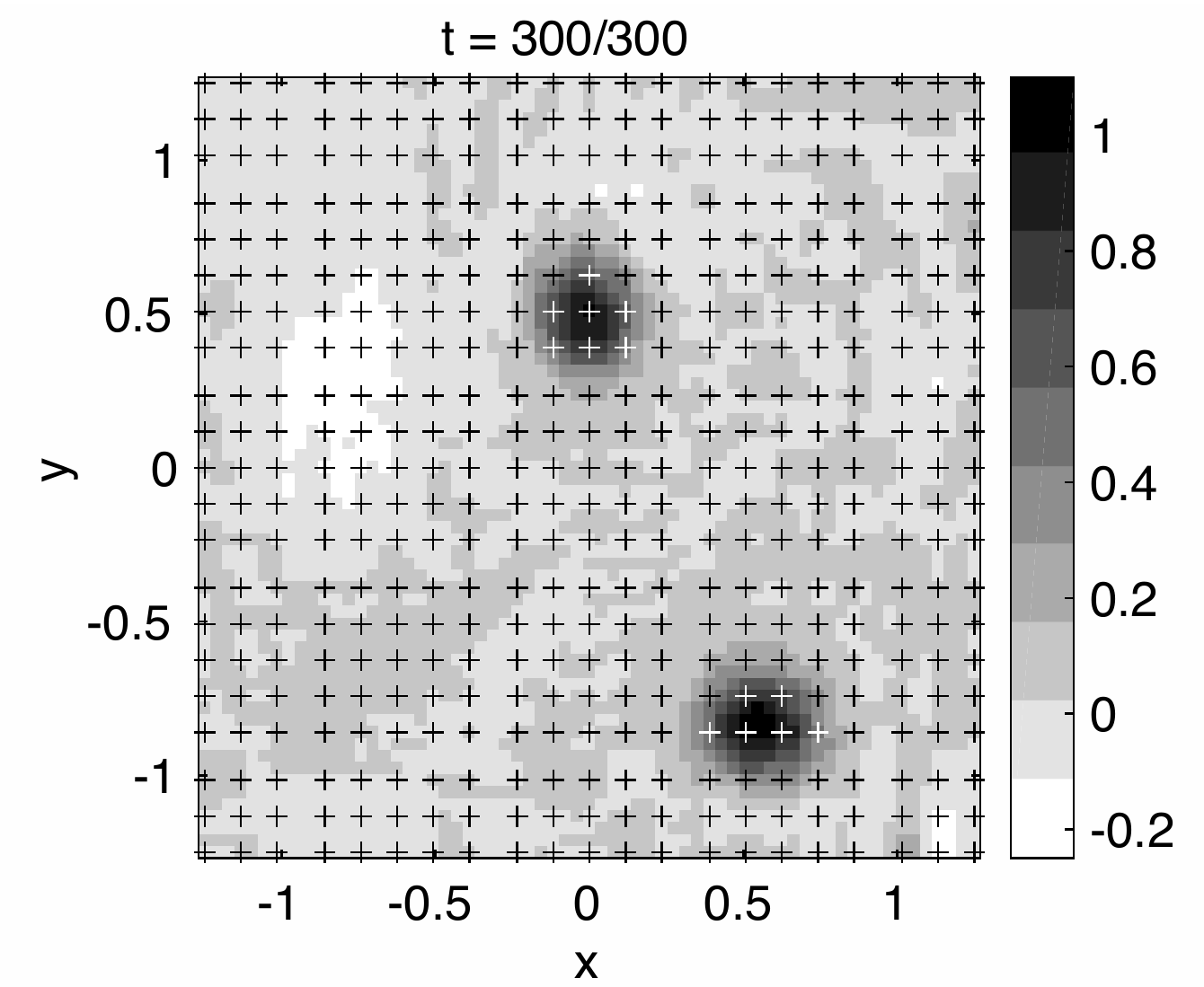}
\end{tabular}
\caption{\textbf{The common vorticity truth trajectory used in all twin experiments.} The array of observation locations for each twin experiment is displaced as black ``+s.'' The initial condition parameters from  \ref{tab:model_params} define two positive vortices. Following the generalized model in  (\ref{eqn:vort_prog_model}), with zero-outflow boundary conditions, the two vortices process counter-clockwise, as in the sequence of panels. The stochastic forcing modulates the rotational speed, separation distance, and centroid location. 
\label{fig:ex_truth_obs}
}
\end{figure}

To confirm the efficacy of strain regularization for penalizing excessive deformations while permiting bulk feature position realignment, a numerical optimization experiment was designed and performed with and without a regularization term (See Figures \ref{fig:strain_reg_expsa}, \ref{fig:strain_reg_expsb}).
 A simple feature modeled after two  coalescing vortices was generated with no background error, and a truth feature was generated by a combination of a rigid transformation and amplitude perturbation. The $L^2$-norm penalty functional from  (\ref{eqn:map_functional}) was used to quantify the residual after realignment by a displacement map $\Phi(z)$.

\subsubsection{Displacement Algorithm Implementation Details}

No basis has yet been prescribed for representing the displacement function $\zeta$. Typical decorrelation lengths for the amplitude perturbations that generate position errors are not global in scale. It is reasonable to assume noise length scales are no greater than the dominant features in the model. For this reason, historically, local parameterizations of displacement functions have been proposed. Mariano \cite{Mari90} decomposed individual contours into angular sections and analyzed displacement along these directions, while Brewster \cite{Brew02} considered a hierarchical model of discretizations, where first three and then one grid step corrections would be sought. Ravela \cite{Rave07} proposed a bicubic spline representation. Frazin \cite{Fraz12}, like Brewster, combined approximation with regularization by considering a hierarchical spline model similar to multigrid techniques, where the displacement function estimate is refined on a sequence of increasingly fine scales. 

There is an issue here regarding how hidden regularizations imposed by the displacement map parameterization affect the model statistics. For any approximation method, the number of discretization points used determines the maximum roughness in the functions which can be modeled. For global methods, this is determined by the highest rank polynomial allowed, while for local methods like splines with fixed rank, the roughness is determined directly by the dimension of the subdomains. The variability of a function can be strongly determined by that of a random displacement map, and local spline interpolants provide a natural way to set the 
 decorrelation length scale of its perturbations. Clearly, a calibration must be performed here to ensure that the map coefficients have the proper statistics for a given amplitude model error, though the details of such calculations do not appear in the published literature. We partially addressed this issue explicitly in Equation (\ref{eqn:pos_covar_proj}), and will take it up again after introducing the particular displacement map parameterization we will use. 

We will adopt the bicubic spline representation, which has a convenient representation as a linear combination of B-splines \cite{Debo62}. With the B-spline formulation rectangular approximations of complex boundaries are easily carried out. For any $p\in C^2(D)$, define a uniform rectangular partition of the domain with grid line intersections at nodes $\{(x_j,y_j)\}$ and spacings $\ell=\Delta x=\Delta y$, and extend this partition one layer of nodes into the exterior of the boundary. For example, a $3\times3$ array of sub-rectangles will have $4\times4 = 16$ nodes in $D$, and a ring of $20$ nodes immediately outside the domain boundary, spaced exactly one grid spacing away. Then $p$ is exactly represented by the linear combination of tensor products of B-splines,
\begin{equation}
p(x,y) = \sum_j a_j B_j(x,y) = \sum_j a_j b(x;x_j) b(y;y_j)
\end{equation}
where the B-spline kernel $b(z;z_j)$ is the piecewise-cubic polynomial
\[
b(z;z_j) = \left\{\begin{array}{cc} 
\frac{1}{6} s^3 & 0\leq s < 1 \\
\frac{1}{6}\left( -3s^3 + 12s^2 - 12s + 4 \right) & 1 \leq s < 2 \\
\frac{1}{6}\left( 3s^3 - 24s^2 + 60s - 44 \right) & 2 \leq s < 3 \\
\frac{1}{6}\left( -s^3 + 12s^2 - 48s + 64 \right) & 3 \leq s \leq 4 \\
0 & \mbox{otherwise} 
\end{array} \right.
\]
and $s = 2+ (z - z_j)/\ell$ is the local normalized coordinate parameterizing the support of the kernel. The extra ring of B-splines around the exterior provides the extra polynomial basis support on each sub-domain along the boundary to make the computation of polynomial coefficients there well-determined. 

As a parameterization for the map function $\Psi$, bicubic B-splines like other linear approximation schemes allow us to conveniently represent the map function as a linear operation on the spline coefficients, $\Psi = B\cdot a$, as well as any linear operation on $\Psi$, such as the displacement map in Equation (\ref{eqn:stream_func_map}), or the local strains induced by the map in Equation (\ref{eqn:eng_strain_tensor}). The position error penalty function in Equation (\ref{eqn:pos_reg_term}) can be represented directly in terms of the spline basis, and amplitude covariance information is mapped onto position perturbation coordinates using the diagnostic relationship between the vorticity and stream functions, Equation (\ref{eqn:poisson_stream}). This resolves the issue of the degree of regularization imposed by the displacement function parameterization, since the position coefficient statistics will be computed directly from the amplitude covariance. For example, if $\delta\omega \sim {\cal N}(0,C_\omega)$, and we want to approximate position uncertainty by a Gaussian distribution, $\delta\Psi\sim {\cal N}(0,C_a$, then we can estimate 
$C_a = T_{a\omega} C_\omega T_{a\omega}^T$. 

For the remainder of this Section, we address a common issue encountered when computing statistical covariances in theoretical, numerical, and observational contexts. Techniques that estimate, propagate, or regularize statistical data often destroy the structure that distinguishes a valid covariance matrix from a generic square matrix.  In these cases, one needs to (re)project the information back into the space of covariance matrices, i.e. symmetric positive semi-definite (SPSD) matrices. The sense in which a candidate projection is closest to the given matrix also must be determined.   Consider a small perturbation to a SPSD matrix $A$ that contributes a negative eigenvalue: take any $u$ in its nullspace and add the rank-1 perturbation $-\epsilon uu^T$ for some $0<\epsilon\ll 1$. Then $A$ is no longer positive semi-definiteness. Then subsequent covariance operations, such as the  Kalman filter covariance update, can yield unexpected results. 
While there are methods of finding the closest SPSD matrix in the 2-norm, they are much more theoretically and computationally complex than in the Frobenius norm. 
A polar factorization of the symmetric matrix component of $A$, given by $B= (A + A^T)/2$, can be accomplished through a singular value decomposition. If $B = U_B\Sigma_B V_B^T$, where $U_B$ and $V_B$ are unitary and $\Sigma_B$ is diagonal and positive semi-definite, then 
\begin{equation}
B = QH= \left( U_B V_B^T \right) \left( V_B \Sigma_B V_B^T \right)
\end{equation}
Observe that the first factor on the right-hand side is  unitary, and since the second factor is a similarity transformation of $\Sigma_B$, then the result  is SPSD.

Anticipating the needs of the two-stage statistical model, we also need an observation operator for the kind of displacement maps posed in  (\ref{eqn:stream_func_map}), or rather the map parameters which index them. These are of course the components, at the observed locations,  of the image of the velocity function under the coordinate transformation $Z = \Phi(z;a)$, for which we can write
\begin{eqnarray}
&& h_p(a; \omega; \{x_{d,j},y_{d,j}\}) =\nonumber \\
&& h_a\left[ \omega \circ \Phi^{-1}(\cdot ;a) ; \{x_{d,j},y_{d,j}\} \right]
\nonumber
\end{eqnarray}
Note that a continuous interpolation for the vorticity function $\omega$ must be used to estimate the velocity at mapped observation locations, for which the most accessible choice is the current forecast $\omega^f$, although other choices may be possible. 

At this point, the only detail left to be addressed before an analysis scheme can be presented relates to the projection of forecast vorticity amplitude information into position coordinates. Such a method already has been introduced in Equation (\ref{eqn:pos_covar_proj}). The new wrinkle here is the imposition of boundary conditions, without which the transformation operator from map parameters to vorticity values, $T_{\omega a}$, will not have full rank. This would preclude the definition of its pseudoinverse $T_{a\omega} = T_{\omega a}^\dagger$, which is the desired projection operator. In other words, the space of displacement map parameters must be restricted to a dimension equal to the degrees of freedom unconstrained by the boundary conditions. There are three sets of conditions to be imposed on the coefficients of the map function $\Psi(z;a) = \sum_{j=1}^{N_c} a_j B_j(z)$, where $N_c = (N_{c,x}+3)\cdot (N_{c,y}+3)$ is the number of B-spline coefficients:
The Laplacian of the map function must be a fixed constant on $\partial D$, so we set $\Delta \Psi = 0$ on these nodes.
The tangent gradient of the map function must be zero on $\partial D$, so we set $\Psi_y = 0$ for $x \in \{x_L, x_U\}$, and $\Psi_x = 0$ for $y \in \{y_L, y_U\}$.
The height of the map function must be fixed on $\partial D$, so we set $\Psi(x_L,y_L) = 0$.
 The tangent condition prevents the map from moving vorticity mass out of the domain $D$ where it is lost to the model. Since the displacement map is defined by derivatives of the map function, any constant term disappears. Constants typically are the first term of any basis for $L^2$ function representations on a compact domain, including bicubic splines, and the height condition removes this degree of freedom. These linear conditions on the map parameters can be expressed by the linear system
\begin{equation}
0 = W\cdot a = (U_W\Sigma_W V_W^T)\cdot a
\end{equation}
where the last equality factors the constraint matrix into its singular value decomposition. If $r_W$ is the rank of the constraint matrix, then the columns of $V_b$, defined to be the last $N_c-r_W$ columns of $V_W$, are a basis for the null space of $W$. Now any set of boundary-constrained map parameters can be represented by $a = V_b\cdot a_b$, and $T_{\omega a_b} = T_{\omega a} V_b$ will have full rank. Then the transformation operator from vorticity model space to boundary-constrained map parameter space is given by
\begin{equation}
T_{a\omega} = V_b T_{\omega a_b}^\dagger
\end{equation}
Given a forecast covariance for the vorticity amplitude perturbations, $P_\omega^f$, then a forecast displacement map parameter covariance, for perturbations subject to zero-outflow boundary conditions, can be estimated from the projection $P_a = T_{a\omega} P_\omega^f T_{a\omega}^T$.

\subsection{Two-stage Analysis Scheme}
At each assimilation time $t_k$ for $1\leq k\leq N_a$ we seek a model (vorticity) state analysis $\omega = \omega^a$ which minimizes a likelihood functional that has been conditioned (or regularized) by a forecast penalty term,
\begin{eqnarray}
\mathcal{J}_a[\omega] &=& \frac{1}{2}\left\| d_V - h_a(\omega; \{z_{d,j}\}) \right\|_{R^{-1}}^2  \nonumber \\
&+& \frac{1}{2}\left\| \omega - \omega^f \right\|_{\left(P_\omega^f\right)^{-1}}^2,
 \label{eqn:ampl_stat_model}
\end{eqnarray}
where $d_V$ are observations of the velocity field at discrete points in $D$.
 Denote the time series of analysis ensembles in enKF by $\mathcal{E}_k^a = \{\omega^a_j(z,t_k)\}$, for $1\leq j\leq N_{ens}$ and $0\leq k\leq N_a$, initialized with the exact initial condition, $\omega_j(z,t_0) = \omega_0(z)$, for each ensemble member. 
 See \ref{enkf}. The goal of the enKF is to compute these analysis ensembles by statistical interpolation of forecast and observation ensembles at each analysis step. Denote this the {\it amplitude analysis step}, which specifically consists of the following: each forecast ensemble $\mathcal{E}_k^f$ is produced by integrating the generalized model over $t\in (t_{k-1},t_k]$, using the previous analysis ensemble $\mathcal{E}_{k-1}^a$ as an initial condition. Each forecast member becomes the estimate of the forecast mean, $\omega^f\to \omega_j^f(\cdot,t_k)$, in their own copy of (\ref{eqn:ampl_stat_model}), with synthetic observations perturbed from original measurements, $d_{V,k,j} = d_{V,k} + \epsilon_{d,j,k}$, with $\epsilon_{d,j,k}\sim {\cal N}(0,R)$, simulating the spread of the same likelihood distribution. The members of each analysis ensemble are correlated by taking the corresponding forecast covariance to be the sample covariance from the forecast ensemble,
\begin{equation}
P_{\omega,k}^f = \frac{1}{N_{ens}-1}\mathcal{E}_k^f \left(\mathcal{E}_k^f\right)^T
\end{equation}
The enKF estimates a Kalman gain matrix, $K_k$, for the entire ensemble such that the linear combination of the simulated forecast and observation ensembles,
\begin{equation} \label{eqn:ampl_anal_update}
\mathcal{E}_k^a = \mathcal{E}_k^f + K_k\cdot \left[ d_{V,k} - h_a\left(\mathcal{E}_k^f; \{z_{d,j}\} \right) \right]
\end{equation}
has minimal analysis variance  when $h_a$ is linear and $N_{ens}\to \infty$. This Kalman gain follows the standard formula, replacing the forecast covariance with the sample covariance of the forecast ensemble,
\begin{equation}
K_k = P_{\omega,k}^f H_{a,k}^T \left( R + H_{a,k} P_{\omega,k} H_{a,k}^T \right)^{-1}
\label{eq:update}
\end{equation}
where $H_{a,k} = \nabla_\omega h_a\left(\omega; \{z_{d,j}\} \right)$ at $\omega = \mathcal{E}_k^f$, the concatenation of linearizations of the vorticity observation operator about each forecast ensemble member. 

The  {\it displacement assimilation} strategy prefaces the amplitude analysis step by a position analysis that preconditions the forecast ensemble. The position statistical model follows the amplitude model in  (\ref{eqn:ampl_stat_model}) in form, although the statistics of the displacement map parameter distribution are regularized following the
(\ref{eqn:strain_covar}) and (\ref{eqn:map_param_reg}). At each analysis step, between the generation of a new forecast ensemble and the assimilation of new observations, we seek a set of displacement parameters $a = a^a$ which optimize the cost functional
\begin{eqnarray} 
\mathcal{J}_p[a] &=& \frac{1}{2}\left\| d_V - h_p(a; \omega^f, \{z_{d,j}\}) \right\|_{R^{-1}}^2 \nonumber \\
&+& \frac{1}{2}\left\| a - a^f \right\|_{\left(P_{a,\alpha}^f\right)^{-1}}^2
\label{eqn:pos_stat_model}
\end{eqnarray}
We note here a few key deviations from the usual enKF definitions: There is no premise for a forecast displacement map, so we take $a^f = 0$ for all ensemble members and at all assimilation times. Consequently, the forecast map ensemble is collapsed, so we must obtain a forecast map parameter covariance some other way. Fortunately, we defined in the previous section the mapping $T_{a\omega}$ with which to project the forecast vorticity amplitude covariance. This information is contained in the forecast vorticity ensemble, and as with the amplitude analysis, to avoid explicitly constructing the forecast vorticity covariance, the forecast increments are first projected into the much lower dimensional map parameter space. Forming the mean vorticity forecast ensemble $\hat{{\cal E}}^f$ from concatenated copies of the ensemble sample mean, then the forecast map parameter covariance can be written as
\begin{eqnarray}
P_{a,k}^f &=& \frac{1}{N_{ens}-1} \left[ T_{a\omega} \left( \mathcal{E}_{k,m}^f - \hat{\mathcal{E}}_{k,m}^f \right)  \right] \cdot  \nonumber \\
&& \left[ T_{a\omega} \left( \mathcal{E}_{k,m}^f - \hat{\mathcal{E}}_{k,m}^f \right)  \right]^T, \nonumber
\end{eqnarray}
and of course still is subject to regularization  before it can be used in  (\ref{eqn:pos_stat_model}). The new index $m$ will be defined momentarily. There is no need to define a forecast map parameter ensemble, since it is 0 by construction. Let the analysis map parameter ensemble be denoted by $\mathcal{A}_{k,m} = \{ a_{j,m}(t_k) \}$.  Since the initial vorticity condition is known exactly, $\mathcal{A}_0 = 0$. At each subsequent analysis time $t_k$, the linear enKF position analysis update can be expressed as
\begin{eqnarray}
\mathcal{A}_{k,m} &=& 0 + L_{k,\alpha}\cdot \left[ d_{V,k} - h_p\left( 0 ; \mathcal{E}_{k,m}^f, \{ z_{d,j} \} \right) \right] \nonumber \\
&=& L_k\cdot \left[ d_{V,k} - h_a\left( \mathcal{E}_{k,m}^f ; \{ z_{d,j} \} \right) \right]
\end{eqnarray}
where the Kalman gain matrix $L_{k,\alpha}$ for the map parameter estimator depends on the strain regularization parameters $\alpha$, and is given by
\begin{equation}
L_{k,\alpha} = P_{a,\alpha,k} H_{p,k}^T \left( R + H_{p,k} P_{a,\alpha,k} H_{p,k}^T \right)^{-1}
\end{equation}
where the linearization of the map parameter observation operator, $H_{p,k}$, must be taken with respect to $a$ and not $\omega$. From the chain rule, the linearization about $\mathcal{A}_{k,m} = 0$ is
\begin{eqnarray}
H_{p,k} 	&&= \nabla_a h_p\left( \mathcal{A}_{k,m} ; \mathcal{E}_{k,m}^f , \{ z_{d,j} \} \right)  \nonumber\\
		&=& \nabla_a \left[ \mathcal{E}_{k,m}^f \circ \Phi^{-1}(\cdot; \mathcal{A}_{k,m} ) \right] \nonumber\\
		&= &\nabla_z \mathcal{E}_{k,m}^f \cdot \nabla_a\Phi^{-1}(\cdot; \mathcal{A}_{k,m} ) 
		 \label{eqn:linearized_pos_op}
\end{eqnarray}
Given the linearity and skew-symmetry of the stream function displacement map $\Phi$ defined in (\ref{eqn:stream_func_map}), then the gradient of the map about $\mathcal{A} = 0$ is
\begin{eqnarray}
&&\left. \nabla_a \Phi^{-1}(\cdot; \mathcal{A} ) \right| \left( \mathcal{A} = 0 \right)  \nonumber \\
&=&  \left. \nabla_a \Phi(\cdot; -\mathcal{A} ) \right| \left( \mathcal{A} = 0 \right)  \nonumber \\
&=& \left[ \begin{array}{c} B_y(\cdot) \\ -B_x(\cdot) \end{array} \right]
\end{eqnarray}
regardless of the map parameters $\mathcal{A}$, where we recall that $B(\cdot)$ are the basis functions of the map function $\Psi$. The resulting map parameter analysis ensemble defines a new forecast ensemble, presumably with less position error near observed features,
\begin{equation}
\mathcal{E}_{k,m+1}^f = \mathcal{E}_{k,m}^f\circ \Phi^{-1}\left( \cdot; \mathcal{A}_{k,m} \right)
\end{equation}
where now can be seen  that $m$ counts the number of times this position correction process is repeated, $1 \leq m \leq M$. 

\section{Vortex Tracking Twin Experiment}

\begin{figure} \centering
\begin{tabular}{rl}
\includegraphics[height = 2in, clip = true, trim = 20 0 30 0]{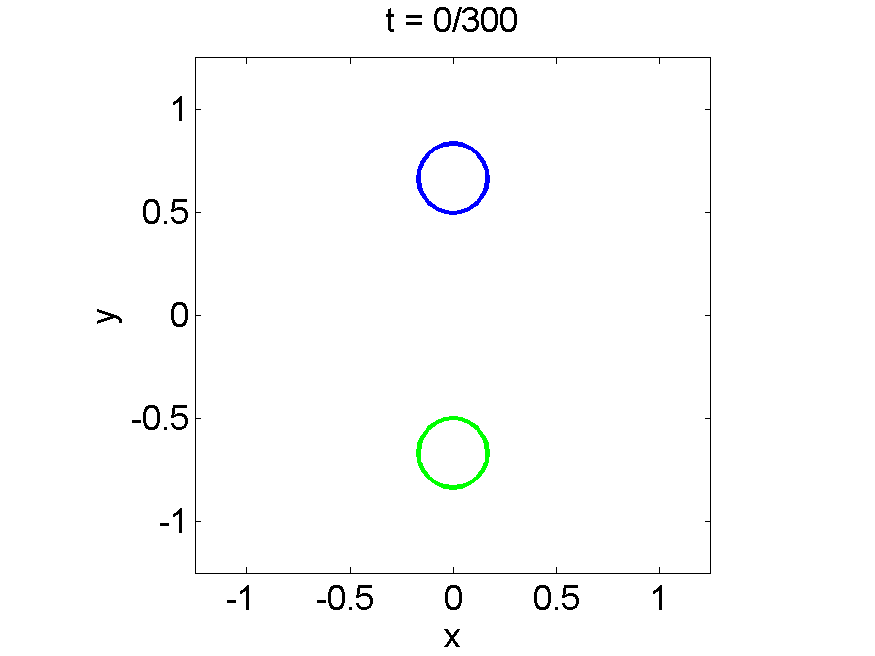} &
\includegraphics[height = 2in, clip = true, trim = 20 0 30 0]{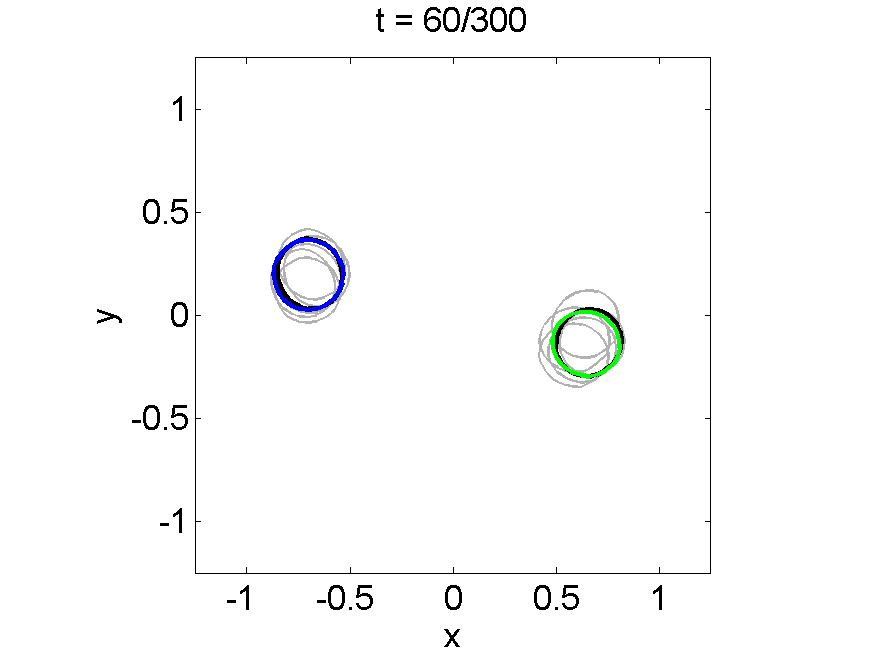} \\
\includegraphics[height = 2in, clip = true, trim = 20 0 30 0]{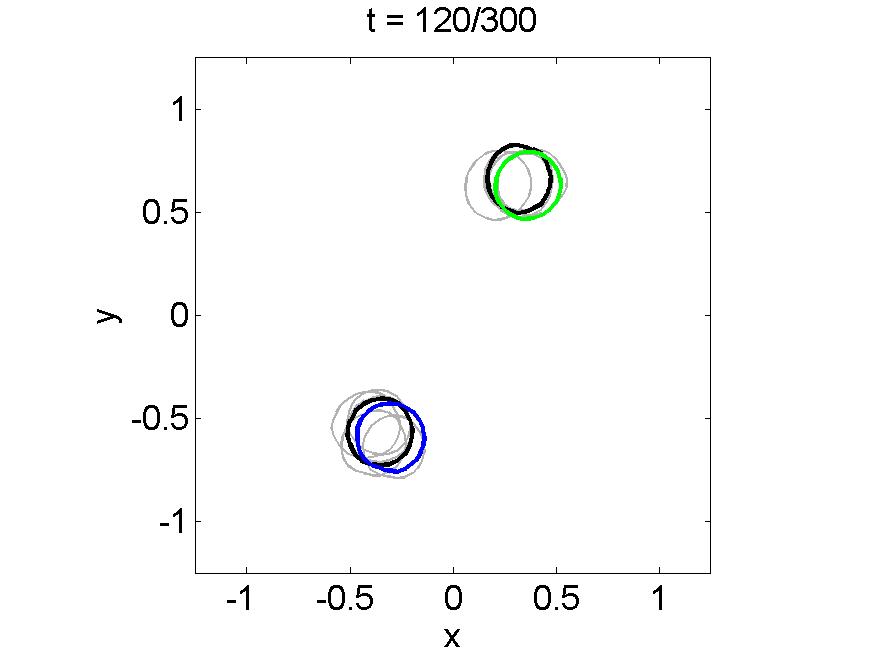} &
\includegraphics[height = 2in, clip = true, trim = 20 0 30 0]{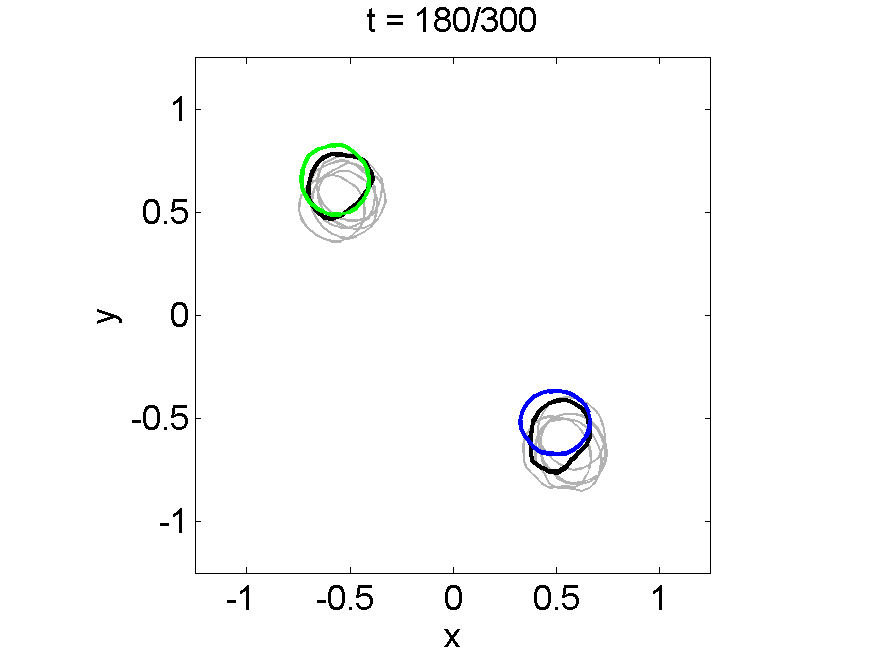} \\
\includegraphics[height = 2in, clip = true, trim = 20 0 30 0]{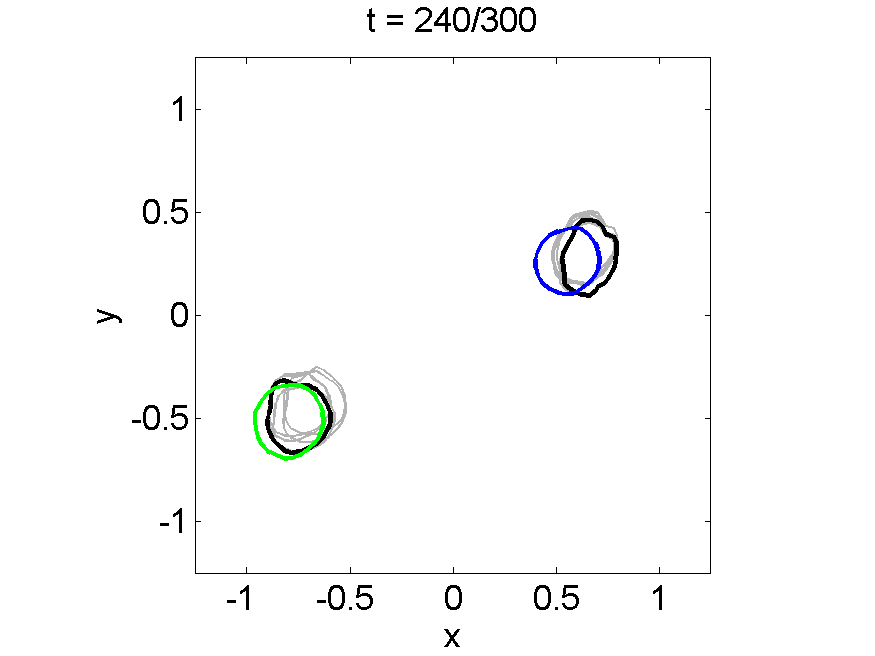} &
\includegraphics[height = 2in, clip = true, trim = 20 0 30 0]{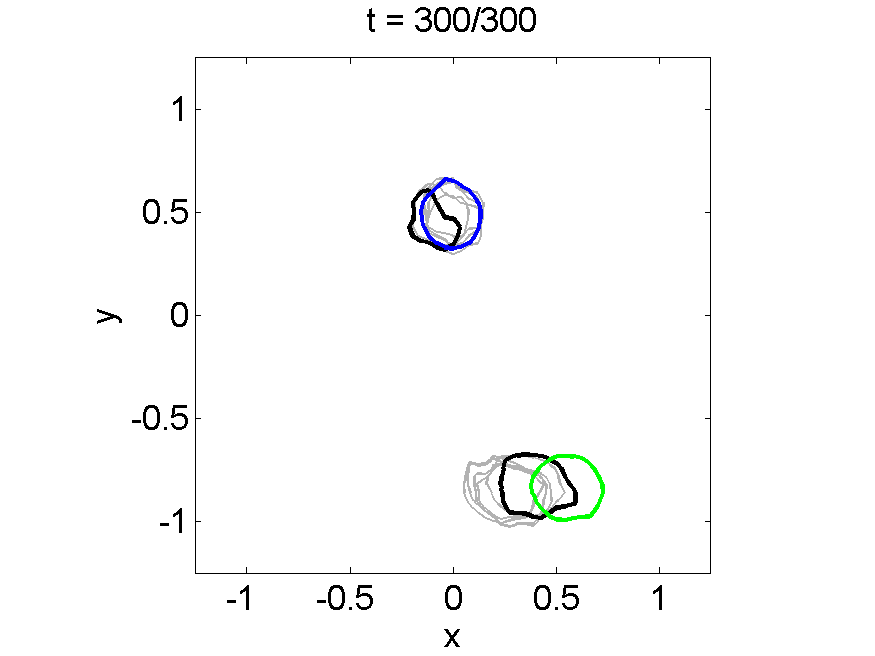}
\end{tabular}
\caption{\textbf{Standard enKF analysis of the true vortex positions in Figure \ref{fig:ex_truth_obs}, with $\mathbf{N_{ens} = 5}$.} The forecast (gray), analysis (black), and truth (blue/green) vortices are depicted by equal-height contours at $\omega = .5$, so as not to contour the underlying noise process. The sparated analysis and truth contours show growing untracked displacement error, as well as significant deformations in the total area and symmetry of the analysis contours. \label{fig:ex_standard}}
\end{figure}

\begin{figure} \centering
\begin{tabular}{rl}
\includegraphics[height = 2in, clip = true, trim = 20 0 30 0]{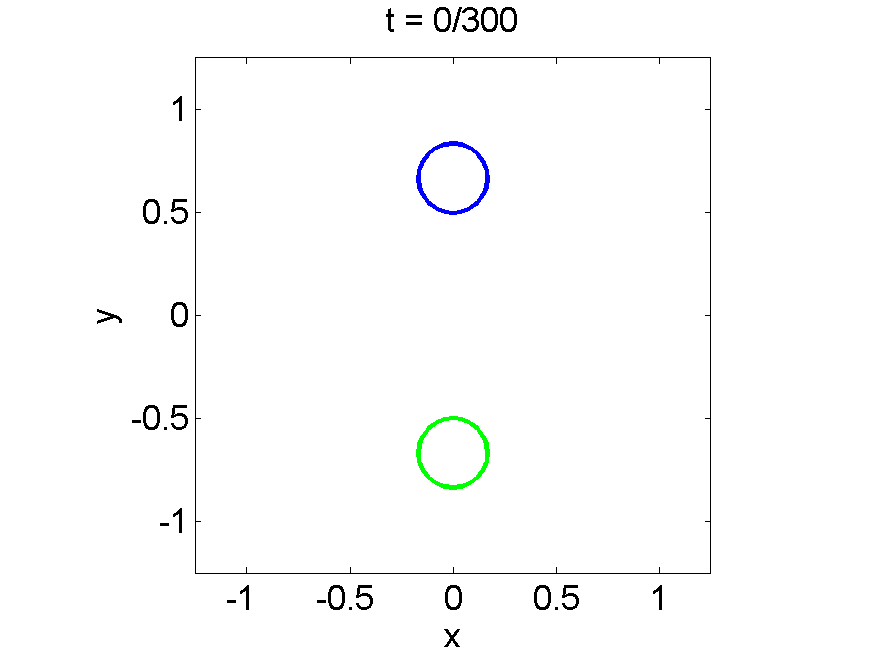} &
\includegraphics[height = 2in, clip = true, trim = 20 0 30 0]{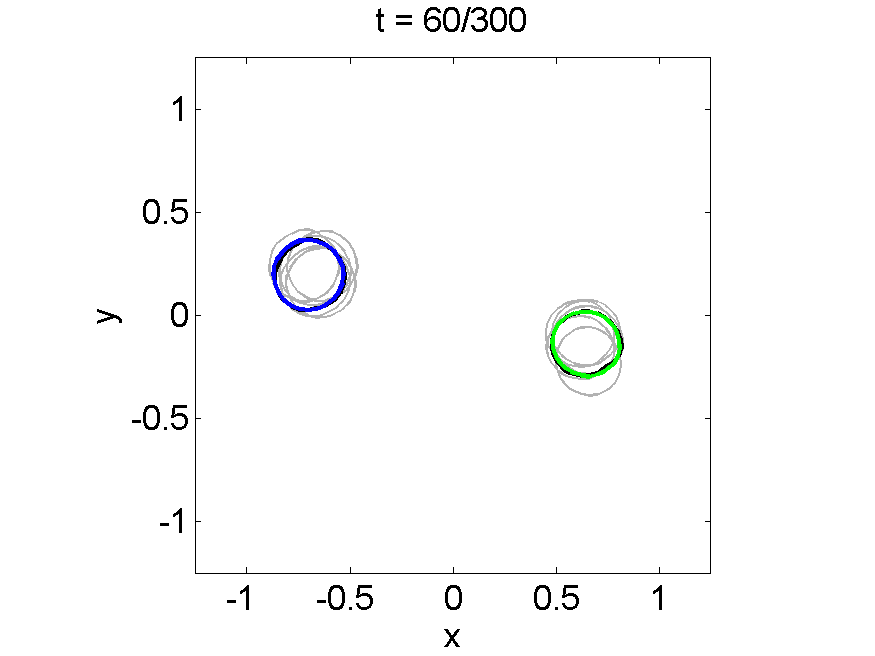} \\
\includegraphics[height = 2in, clip = true, trim = 20 0 30 0]{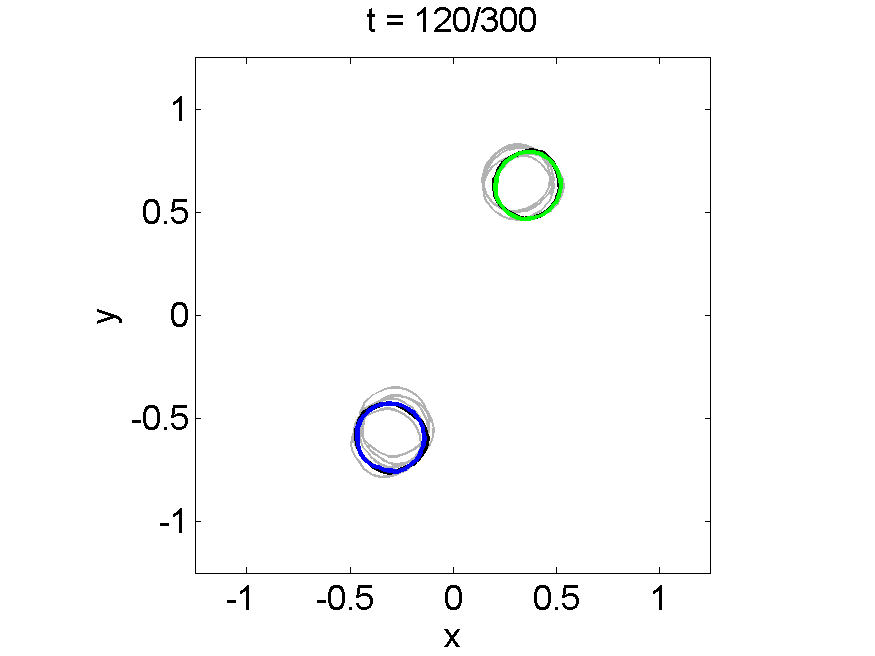} &
\includegraphics[height = 2in, clip = true, trim = 20 0 30 0]{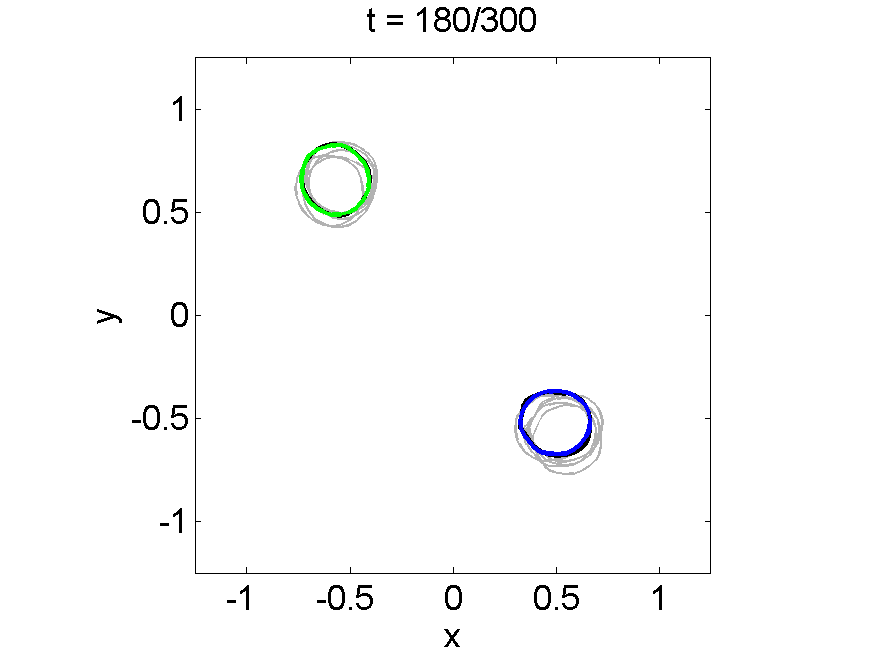} \\
\includegraphics[height = 2in, clip = true, trim = 20 0 30 0]{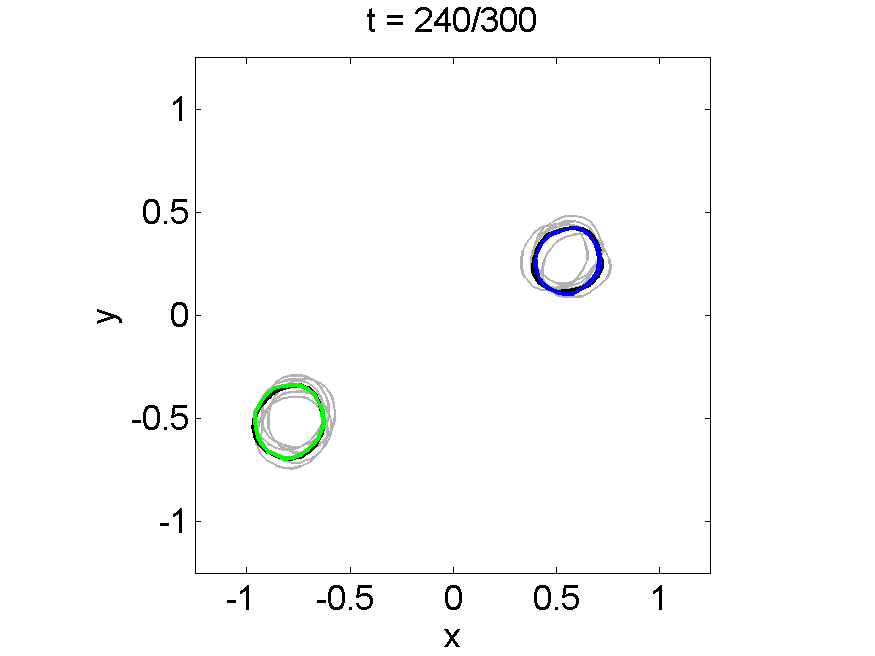} &
\includegraphics[height = 2in, clip = true, trim = 20 0 30 0]{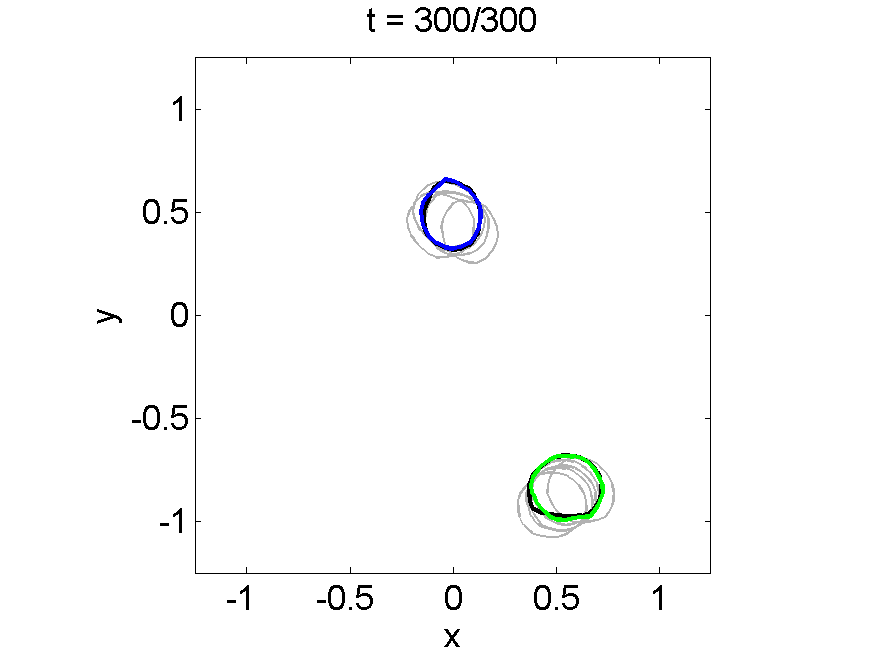}
\end{tabular}
\caption{\textbf{Displacement enKF analysis of the true vortex positions in Figure \ref{fig:ex_truth_obs}, with $\mathbf{N_{ens} = 5}$.}. Same conditions as in Figure \ref{fig:ex_standard}. The truth and analysis observations coincide throughout the simulation, and little deformation or change in the total area of the contour can be observed. This indicates that displacement correction is helping to maintain feature structure and reduce analysis error variance. \label{fig:ex_position}}
\end{figure}

\begin{figure} \centering
\includegraphics[height = 2in, clip = true, trim = 30 0 50 0]{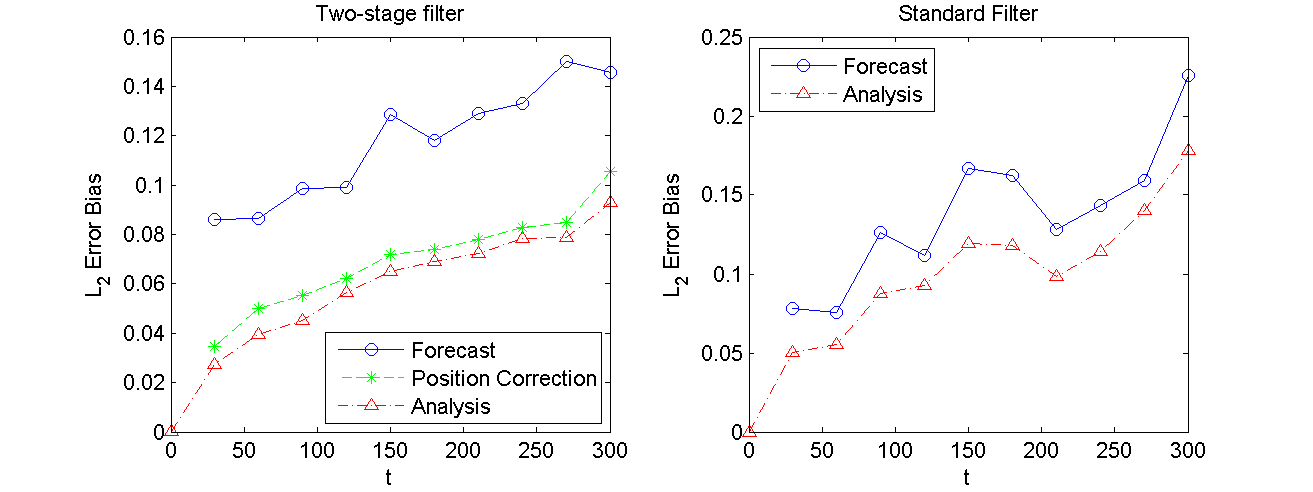} \\
\includegraphics[height = 2in, clip = true, trim = 30 0 50 0]{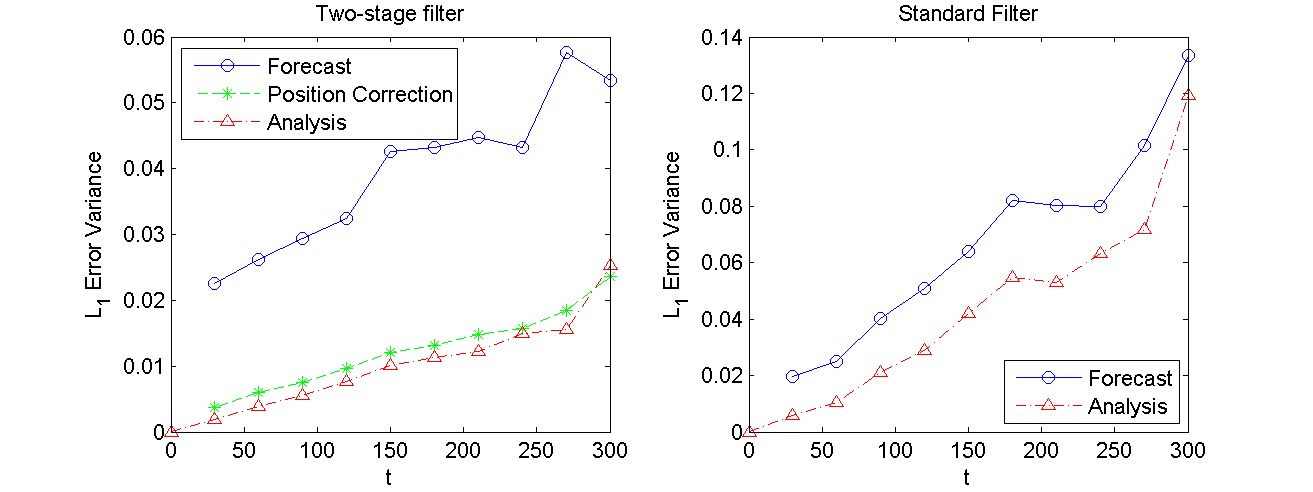}
\caption{\textbf{Time sequences of assimilation error biases (top) and variances (bottom).} These are averaged over all $N_{ens} = 5$ runs of the standard (right) and two-stage (left) enKF. This averaging is intented to smooth out the differences in observation and synthetic observation errors, and ensemble perturbations. Most of the forecast bias is removed by the displacement correction, with smaller gains from the subsequent model state analysis. }
 \label{fig:sample_error_stats}
\end{figure}

A twin experiment was used to measure forecast and analysis errors by comparing the vorticity trajectory predicted by the standard and the displacement enKF to a truth which, like the forecast ensembles, is a simulation of the stochastic vorticity model. Observations of the entire truth are not permitted to be used with the filter, only a grid of noisy velocity measurements that have been synthetically generated from the true vorticity trajectory. The analysis times were spaced far enough apart to begin to see performance differences between the standard and the displacement enKF. Constant-height contours at assimilation times are provided in Figures \ref{fig:ex_standard} and \ref{fig:ex_position} to illustrate the effectiveness of each enKF in forecasting and tracking the truth. The number of ensemble members was $N_{ens} = 5$. A more general argument comparing the efficiency and accuracy of each filter is made in  Section \ref{performance}. The contours show that the strain regularized displacement maps appear to generate tighter forecasts. The contours of the two-stage analyses preserve the circular feature geometry of the truth with much higher fidelity than the analyses provided by the standard filter. The loss of structure may be due to the bias in forecast mean and inflation of forecast uncertainty that can arise from a spatially dispersed set of predictions of the same strong feature, such as an ensemble of realizations of the same vortex. Details supporting this claim will be presented in  Section \ref{performance}. That most of the forecast displacement is corrected without these distortions in the two-stage analyses is evidence that strain regularization is penalizing excessively tortuous displacement maps. 
In addition to the preservation of features, the performance of the standard and displacement enKF will be evaluated and compared by quantifying the resulting distributions of forecast error, $\epsilon^f := \omega^t - \omega^f$, and analysis error, $\epsilon^a := \omega^t - \omega^a$. The error bias, $\mathbb{E}(\epsilon)$, in the forecast and analysis of each filter is measured by comparing these to the true vorticity function at each analysis time in the $L_2$-norm. The error variance, Var$(\epsilon)$,  is measured by computing the sample variance of these errors over several repetitions of the twin experiment, and measuring this variance in the $L_1$-norm. These are taken to be functions on the entire model domain, $D$, and not just the observed subset, $D_d\subset D$; since the model interpolates the data, it is expected that the most accurate predictions will occur at observed locations. When the domain is only sparsely observed, the errors measured on $D_d$ are not indicative of the accuracy of predictions made on all of $D$. Figure \ref{fig:sample_error_stats} gives the time sequences of forecast and analysis error biases and variances for the standard and two-stage filters. More specifically, these are the average error bias and variance statistics for the 16 repetitions of the twin experiment with $N_{ens} = 5$. The consistently increasing error statistics are due to the fact that the boundary conditions do not permit energy to leave the system, as the stochastic forcing in the vorticity model continually increases the model variance over time. It was shown in Figure 1 that the model uncertainty is dominated by position perturbations in the vortex positions, as opposed to amplitude perturbations. Likewise, the majority of forecast error removal is done through the position analysis. The standard filter does not define position error directly, so it cannot remove as much of the forecast variance with each analysis step. This is why the  error bias and variance is consistently higher with the standard enKF than for the displacment enKF.

\subsection{Performance Comparison}
\label{performance}
\begin{figure} \centering
\includegraphics[height = 2in, clip = true, trim = 30 0 50 0]{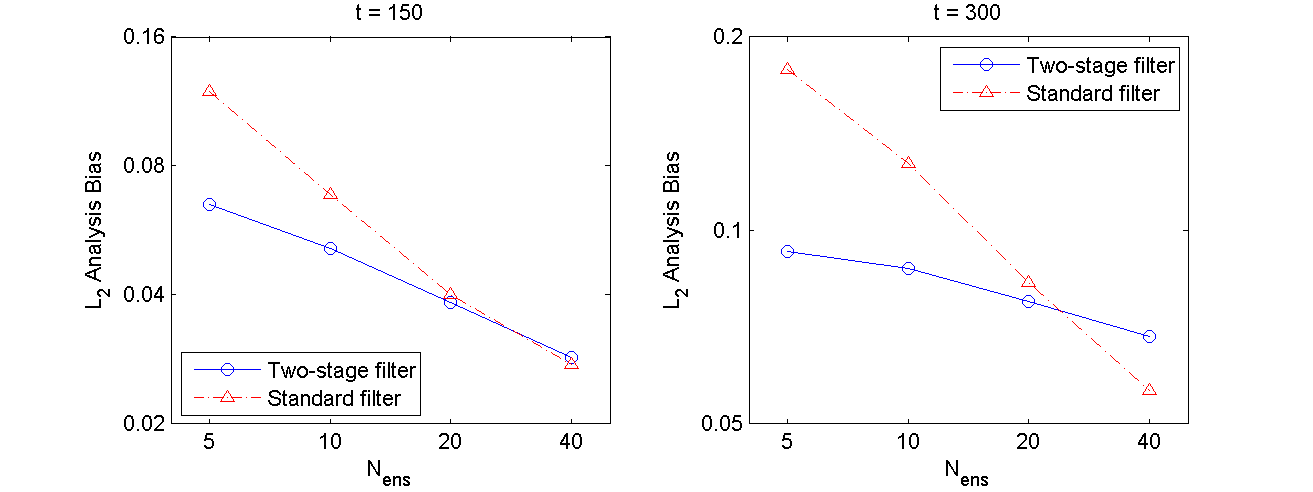} \\
\includegraphics[height = 2in, clip = true, trim = 30 0 50 0]{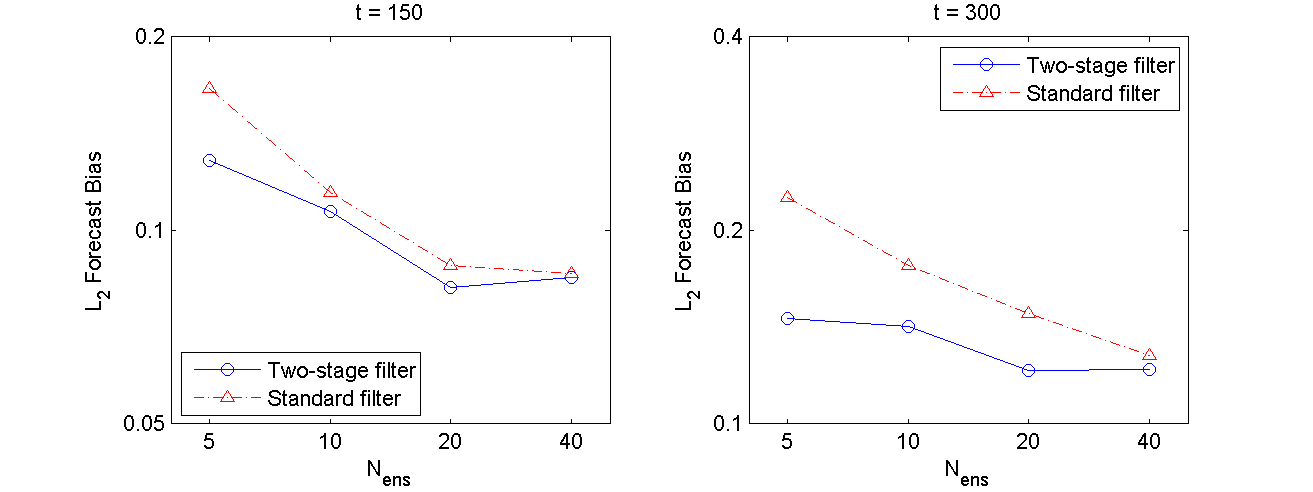}
\caption{\textbf{Decay of forecast and analysis error biases, $L_2 \mathbb{E}(\epsilon)$, with increasing ensemble size.} The results are given at two model times, $t = 150$ and $t = 300$, where the increase in model error variance with time simulates a sensitivity analysis. (top panels) The analysis error bias trends toward  convergence for both filters, but at a slower rate for the displacement filter. A smaller proportionality constant allows the displacement filter to out-perform the standard one for smaller ensembles, and for ensembles exceeding approximately 30 members, the standard enKF better constrains analysis error bias. (bottom panels) The forecast error bias need not converge, even if the analysis error bias does. For the ensemble sizes tested, the displacement filter provides consistently better predictions, though the results suggest this also may only be true for smaller ensembles. The performance distinction widens as model error variance increases.}
 \label{fig:vortex_bias_error}
\end{figure}

The twin experiment of the previous section was repeated several times for several different ensemble sizes, $N_{ens}$ (Table \ref{tab:twin_reps}). Only the true vortex trajectory was was reused for each repetition, which facilitates a comparison of the performance of each enKF version, while controlling for variations in observation noise and ensemble selection. For each experiment repetition, the observation errors and synthetic observations, as well as the model perturbations during ensemble integration, are driven by independent random processes. This variation is removed by averaging the error bias and variance statistics of all trials with a given ensemble size. Sensitivity to observation noise is not performed here. Rather, since the model variance naturally increases with time, then an examination of the forecast and analysis errors at different model times, $t = 150$ and $300$, provides sensitivity information with respect to changes in a model variance parameter. 
\begin{table}[t] 
\begin{center}
\caption{\textbf{Repetitions of twin experiment with stochastic vorticity model.} \label{tab:twin_reps}}
\begin{tabular}{|l|cccc|}
\hline
$N_{ens}$	& 5	& 10	& 20	& 40	\\ \hline
Sample size	& 16	& 12	& 10	& 8	\\ \hline
\end{tabular}
\end{center}
\end{table}
 Figure \ref{fig:vortex_bias_error} provides numerical evidence that both the standard and the displacement enKF estimates converge, although the displacement enKF does so more slowly. While the analysis bias error left by the standard enKF decays with order approximately $2/3$ and is robust to changes in model variance, the analysis bias of the displacement enKF is approximately half that, or order $1/3$, and slows by half again to order $1/6$ as model variance increases. 

The forecast error bias indicates the predictive accuracy of the filter. It need not decay to zero, even if the analysis error does, since the accuracy of the prediction depends on more than the initial condition. Hence, a meaningful convergence rate is difficult to define or compute. However, a qualitative comparison of the forecast biases of the two filters shows the 
displacement  filter is helping to improve predictive accuracy. The gains are nominal for low model variance, but as it increases, the forecast bias of the standard enKF increases for smaller ensembles ($N_{ens}<40$), while the performance of the
displacement enKF remains stable. This means the relative accuracy of the predictions of the displacement filter improves over that of the standard filter as model error increases. 

\begin{figure} \centering
\includegraphics[height = 2in, clip = true, trim = 30 0 50 0]{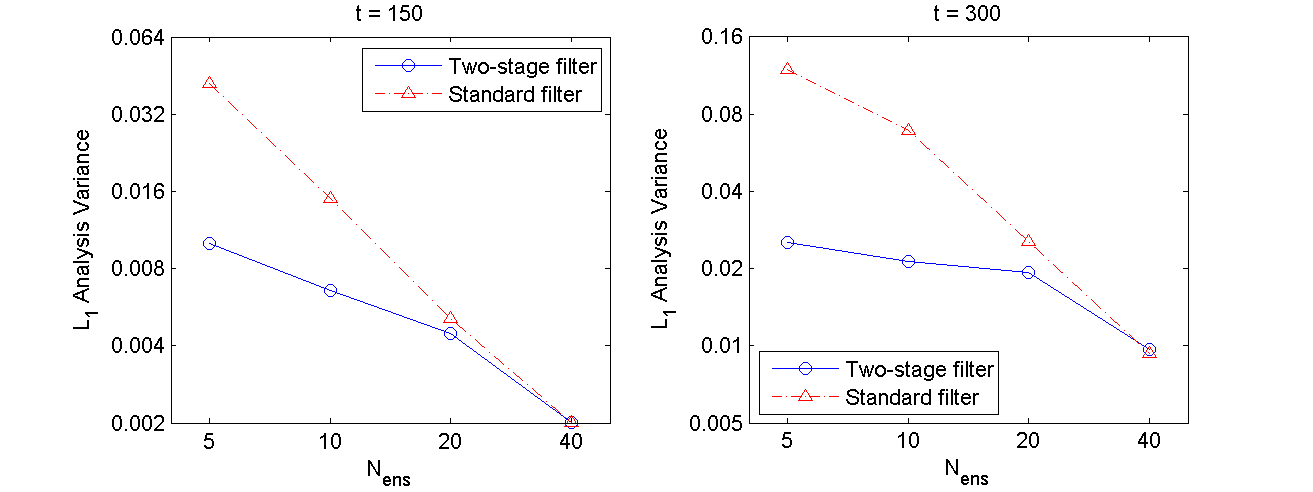} \\
\includegraphics[height = 2in, clip = true, trim = 30 0 50 0]{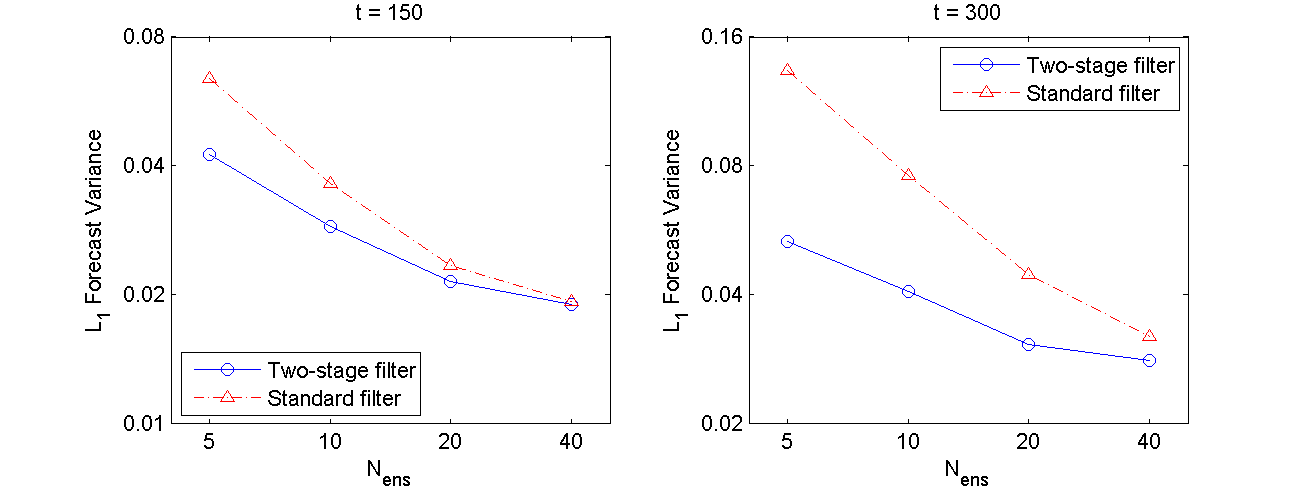}
\caption{\textbf{Decay of forecast and analysis error variance, $L_1$ Var$(\epsilon)$, with increasing ensemble size.} The results are given at two model times, $t = 150$ and $t = 300$, where the increase in model error variance with time simulates a sensitivity analysis. The displacement enKF provides substantial and widening improvements in both the analysis (top) and forecast (bottom) error variance, over that of the standard enKF, as the model error increases. The reduction in error variance increases with smaller ensembles, and unlike the error bias, there is no clear change in the performance leader at larger ensembles. The behavior at larger ensembles suggests the displacement filter converges with the performance of the standard filter; then the displacement method would be a safe choice to improve forecast predictions, with attractive performance in undersampled conditions. }
\label{fig:vortex_var_error}
\end{figure}

The forecast and analysis error variance measures the confidence one can have in the state estimates provided by the filter, before and after, respectively, the assimilation of new data. Measurements of the uncertainty in predictions are as important as the predictions themselves, and in these terms the quantitative evidence in support of using a forecast displacement  preconditioner is positive (Figure \ref{fig:vortex_var_error}). The analysis error variance of the standard enKF decays at a robust order $4/3$, and again for smaller ensembles the modified enKF is slower to converge. However, for medium-sized ensembles and possibly larger, the displacement filter appears to regain the convergence of the standard method, so that for all values tested, the displacement enKF provides  more accurate analyses. 

The displacement enKF also provides greater certainty in its state predictions, based on an analysis of the improvements in forecast variance reduction (Figure \ref{fig:vortex_var_error}). For the small- to medium-sized ensembles tested, the forecast error variance from the displacement filter was equal to or reduced compared to that of the standard filter. As the ensemble size decreases, the variance reduction increases up to 33\% of the standard filter forecast error variance (Table \ref{tab:vortex_comp_efforta}). The improvement in variance reduction is more pronounced as model variance increases, in which the percentage decrease in forecast error variance ranges from 12\% to 60\% depending on the ensemble size. Another way of looking at this data is in terms of the computational effort that is saved using the displacement filter to achieve  a prescribed  level of forecast error variance (Tables \ref{tab:vortex_comp_efforta} and \ref{tab:vortex_comp_effortb}). As the model variance increases, the percentage reduction in the required ensemble size moves from approximately 30\% to about 60\%. Either measured in terms of forecast variance reduction or computational efficiency, the evidence demonstrates that displacement  filter can improve vortex tracking.
\begin{table}[t] 
\begin{center}
\caption{\textbf{Percentage computational savings by using displacement enKF,  at $t=150$.
The $L_1$ estimates are for the forecast variance, the last row is the variance reduction percentage. }} \label{tab:vortex_comp_efforta}
\begin{tabular}{|ccccc|}
\hline	
 $N_{ens}$						& 5		& 10		& 20		& 40	    \\ \hline
 $L_1$  (standard)		& .064	& .036	& .023	& .019    \\ \hline
 $L_1$ (two-stage)		& .043	& .029	& .021	& .019    \\ \hline
 \bf \%  reduction			& \bf 33	& \bf 19	& \bf 9	& \bf 0    \\ \hline
 \end{tabular}
\end{center}
\end{table}

\begin{table}[t] 
\begin{center}
\caption{\textbf{Percentage computational savings by using a displacement enKF, at $t=300$.
The $L_1$ estimates are for the forecast variance, the last row is the variance reduction percentage.}} \label{tab:vortex_comp_effortb}
\begin{tabular}{|ccccc|}
\hline
 $N_{ens}$						& 5 		& 10 		& 20 		& 40 	    \\ \hline
 $L_1$  (standard)		& .133	& .076	& .045	& .032    \\ \hline
 $L_1$  (two-stage)		& .053	& .041	& .031	& .028    \\ \hline
 \bf \%  reduction			& \bf 60	& \bf 46	& \bf 31	& \bf 12  \\ \hline
\end{tabular}
\end{center}
\end{table}

\begin{table}[t] 
\begin{center}
\caption{\textbf{Percentage computational savings by using the displacement enKF, at $t=150$.}}
\label{tab:vortex_comp_savingsa}
\begin{tabular}{|cccc|}
\hline
 Desired $L_1$ forecast variance		& .02		& .03		& .04		\\ \hline
 Required $N_{ens}$ (standard)		& 35		& 14		& 9 		\\ \hline
 Required $N_{ens}$ (two-stage)		& 30		& 9		& 6		\\ \hline
 \bf \% reduction in computation		& \bf 14	& \bf 36	& \bf 33	\\ \hline
\end{tabular}
\end{center}
\end{table}

\begin{table}[t] 
\begin{center}
\caption{\textbf{Percentage computational savings by using the displacement enKF, at $t=300$.}}
\label{tab:vortex_comp_savingsb}
\begin{tabular}{|cccc|}
\hline
 Desired $L_1$ forecast variance		& .032	& .04		& .05 		\\ \hline
 Required $N_{ens}$ (standard)		& 40		& 25		& 18		\\ \hline
 Required $N_{ens}$ (two-stage)		& 18		& 11		& 6		\\ \hline
 \bf \% reduction in computation		& \bf 55	& \bf 56	& \bf 67	\\ \hline
\end{tabular}
\end{center}
\end{table}

\section{Conclusions}

We have proposed a data assimilation strategy that we call displacement assimilation. It introduces an extra stage in the traditional sequential data assimilation strategy, wherein a kinematically constrained transformation is applied in order to preserve geometric features of the 
state vector. Preservation of features in estimates may be essential to pinpointing sources accurately, or to estimating tracks of waves, storm systems and tracers whose transport is dominated by advection. While the methodology is not new, we have introduced new ideas that are particularly applicable to waves and flows in the ocean; we also suggest a regularization procedure that might have an impact on other strategies to track/correct features in estimates.
 If the statistics of the forecast distribution are represented by the ensemble, then conditioning the ensemble is a type of change of measure on the forecast distribution. In particular, this transformation steers some forecast ensemble members closer to observations wherever strong features are observed. 
 
In the numerical simulations we were able to demonstrate that displacement assimilation permitted the retention of structure throughout the data assimilation procedure, and further, reduced analysis error variance. Within the context of ensemble data assimilation  the effectiveness of displacement assimilation was more dramatic for small ensembles. 
When model error was increased, the displacement assimilation required fewer ensemble members in the estimation process than the standard enKF. These two outcomes bode well in using the proposed scheme in data assimilation involving large scale model simulations in which computing many ensemble members is cost-prohibitive.
Displacement assimilation does not have to be applied to the enKF exclusively; it can be invoked in any other sequential estimation method.
 The displacement correction adds
computational cost but in the vortex problem this was not an issue because local splining of the map function proved effective. This type of analysis  localization might prove useful in other dynamic problems.

\section*{Acknowledgements}
This work was supported by GoMRI and by NSF DMS  grants   0304890, and 	NSF OCE grant 1434198.
JMR wishes to thank Stockholm University, where some of this work was done, and their  Rossby Fellowship program.

\appendix

\section{The Ensemble Kalman Filter}
\label{enkf}

The ensemble Kalman Filter (enKF) \cite{Ev92,Even09} is an ensemble-based data assimilation technique for sequential problems. 
As in the standard Kalman Filter, the filter completes the assimilation process in two stages: a forecast and an analysis stage. Unlike the linearized Kalman Filter, the extended Kalman Filter (EKF) (see \cite{Jazw70} for background on the KF and EKF). The enKF uses an ensemble of model runs to compute a mean proposal at the next filtering step.  the analysis step is the same as in the Kalman Filter case. Namely, a Gaussian approximation is made of the local posterior density, of the state vector given observations. With this assumption it is possible to write down the update on the mean, given observations, and an estimate of the posterior covariance. In the process the ensemble (model) mean, and a sample approximation of the (model) covariance are used in 
calculating the update and the Kalman gain. The enKF is particularly attractive because it is easy to code and requires minimal modifications to existing codes representing models.

At time $t_0$, the random perturbations of  initial conditions are $y_j^a(t_0) \sim p(y_0)$ for each $1\leq j\leq N_{ens}$. Expressing ensembles as matrices of concatenated ensemble members, so that each analysis ensemble for $0\leq k\leq N$ and forecast ensemble for $1\leq k\leq N$ can be written as the matrices
\[
{\cal E}_k^a = \left[ y_1^a(t_k) | ... | y_{N_{ens}}^a(t_k) \right]  \quad  {\cal E}_k^f = \left[ y_1^f(t_k) | ... | y_{N_{ens}}^f(t_k) \right]
\]
Then each forecast ensemble ${\cal E}_k^f$ at time $t_k$ is just the column-wise integration of the model equations, using the previous analysis ensemble ${\cal E}_{k-1}^a$ at time $t_{k-1}$ as initial conditions. We will also represent the sample mean of an ensemble as  
\[
\hat{{\cal E}}_k^f = \left[ \hat{y}_k^f | ... | \hat{y}_k^f \right],
\]
where $\hat{y}_k = N_{ens}^{-1}\sum_{j=1}^{N_{ens}} y_j(t_k)$.
The  ensemble forecast covariance is
\[
P_k^f = \frac{1}{N_{ens}-1} \left( {\cal E}_k^f - \hat{{\cal E}}_k^f \right) \left( {\cal E}_k^f - \hat{{\cal E}}_k^f \right)^T.
\]
 Let the ensemble of synthetic observation realizations be denoted by ${\cal D}_k = \left[ d_k + \epsilon_{k,1} | ... | d_k + \epsilon_{k,N_{ens}} \right]$, where each normal variate $\epsilon_{k,j}\sim \cal{N}(0,R)$. The ensemble analysis and the forecast are related to each other  via the standard Kalman filter estimator: The linear Kalman update for the analysis ensemble is given by
\begin{equation} 
{\cal E}_k^a = {\cal E}_k^f + K_k\cdot \left[ {\cal D}_k - h\left({\cal E}_k^f\right) \right]
\label{eqn:enkf_analysis}
\end{equation}
where the observation operator $h(\cdot)$ is applied column-wise to each of the forecast ensemble members. The shared Kalman gain matrix can be expressed as
\begin{equation}
K_k = P_k^f H_k^T \left( H_k P_k^f H_k^T + R \right)^{-1}
 \label{eqn:enkf_gain_matrix}
\end{equation}
and $H_k = \left. \nabla_y h({\cal E}) \right| ( {\cal E} = {\cal E}_k^f )$ denotes the column-wise gradient of the ensemble observation operator evaluated at each forecast ensemble member. 

When the observations are sparse so that $N_d\ll N$, a more efficient variant of  (\ref{eqn:enkf_gain_matrix}), in terms of the observed forecast ensemble $h({\cal E}_k^f) \in \Omega_d$, is used:
\[
K_k = \frac{{\cal E}_k^f h({\cal E}_k^f)^T }{N_{ens}-1} \left[ \frac{ h({\cal E}_k^f) h({\cal E}_k^f)^T }{N_{ens}-1} + R \right]^{-1}
\]
One still can use the linearized observation matrix, $h({\cal E}_k^f) \equiv H_k {\cal E}_k^f$, and achieve the same gain in efficiency. With this representation of the enKF analysis in  (\ref{eqn:enkf_analysis}), it can be written in as a weighted combination of the ensemble members,
\[
{\cal E}_k^a = {\cal E}_k^f \cdot W_k\left[ {\cal D}_k; h\left( {\cal E}_k^f \right) , R \right].
\]
Now, the weights $W_k$, rather than being dependent on the likelihood functional, are dependent on the Gaussian parameters of the likelihood distribution. 
\section{Computational Parameters}
\label{tab:model_params}
Parameters of the generalized vorticity model and 
parameters of the two-stage filter: \\
\begin{tabular}{l}
Domain geometry: $D = \left[ -1.25 , 1.25 \right]^2$ \\
Domain discretization: $N_x = N_y = 64$ ($N = 4225$) \\
Initial condition parameters: \\
 Vortex centroids  $x$:  $x_{c,1} = x_{c,2} = 0$ \\
 Vortex centroids  $y$:  $y_{c,1} = -y_{c,2} = 2/3$ \\
 Vortex radii:  $r_{s,1} = r_{s,2} = 1/3$ \\
 Vortex amplitudes: $a_1 = a_2 = 1$ \\
Time integration interval: $t\in [0,300]$ \\
Integration time step: $\Delta t = 0.05$ \\
Model error pointwise standard deviation: $\sigma_V = 10^{-3}$ \\
Model error decorrelation length: $r_V = 0.707$ \\
Model error mode eigenvalue tolerance: $\lambda(Q_V) \geq 10^{-14}$ \\
Model error boundary dampening width: $r_b = .1$ \\
Position analysis iterations: $M = 3$ \\
Map parameter discretization: $N_{c,x} = N_{c,y} = 20$ ($N_c = 441$) \\
Regularization parameters: $\alpha_n = \alpha_s = 50$ \\
Observation dimensions: $N_{d,x} = N_{d,y} = 20$ ($N_d = 441$) \\
Observation error standard deviation: $\tau = 10^{-3}$ \\
Model error pointwise standard deviation: $\sigma_V = 10^{-3}$ \\
Model error decorrelation length: $r_V = 0.707$ \\
Model error mode eigenvalue tolerance: $\lambda(Q_V) \geq 10^{-14}$ \\
Model error boundary dampening width: $r_b = .1$ \\
Assimilation time step: $\Delta t_a = 30$ \\
\end{tabular}

\section*{References}

\bibliography{doubwell,uq}

\end{document}